\documentclass[prd,nofootinbib,twocolumn,preprintnumbers]{revtex4}
\pdfoutput=1

\usepackage{float}
\usepackage{color}
\usepackage{xcolor}
\usepackage[utf8]{inputenc}
\usepackage[squaren]{SIunits}
\usepackage{amsmath}
\usepackage{amsfonts}
\usepackage{slashed}
\usepackage{graphicx}
\usepackage{subfigure}
\usepackage{rotating}
\usepackage{enumitem}  
\usepackage{pifont}
\usepackage{amssymb,epsfig,verbatim,mathrsfs,array,layout,textcomp,latexsym}
\usepackage{multirow}

\allowdisplaybreaks[1]

\renewcommand{\{}{\left\lbrace}
\renewcommand{\}}{\right\rbrace}
\renewcommand{\[}{\left\lbrack}

\renewcommand{\Re}[1]{\mathrm{Re}\!\{#1\}}

\newcommand{\kil}{\,\text{k}}
\newcommand{\mill}{\,\text{M}}

\def \trace{\text{Tr}}

\begin{document}

\preprint{DO-TH 20/10}

\title{Rare charm $\boldsymbol{c\to u\,\nu\bar{\nu}}$ dineutrino null tests for $\boldsymbol{e^+e^-}$--machines}

\author{Rigo Bause}
\email{rigo.bause@tu-dortmund.de}
\author{Hector Gisbert}
\email{hector.gisbert@tu-dortmund.de}
\author{Marcel Golz}
\email{marcel.golz@tu-dortmund.de}
\author{Gudrun Hiller}
\email{ghiller@physik.uni-dortmund.de}
\affiliation{Fakultät für Physik, TU Dortmund, Otto-Hahn-Str.\,4, D-44221 Dortmund, Germany}

\begin{abstract} 
Rare $|\Delta c|=|\Delta u|=1$ transitions into dineutrinos are strongly GIM-suppressed and constitute excellent null tests of the standard model.
While branching ratios of $D \to P \,\nu \bar \nu$, $D \to P^+ P^- \, \nu \bar \nu$, $P=\pi, K$, baryonic $\Lambda_c^+ \to p \,\nu \bar \nu$, and 
$\Xi_c^+ \to \Sigma^+ \, \nu \bar \nu$ and inclusive $D \to X  \nu \bar \nu$ decays are experimentally unconstrained, signals of new physics can be just around the corner.
We provide model-independent upper limits on branching ratios reaching few $\times 10^{-5}$ in the most general case of arbitrary lepton flavor structure, $\sim 10^{-5}$ for scenarios with charged lepton conservation and few $\times 10^{-6}$ assuming lepton universality. 
We also give upper limits in $Z^\prime$ and leptoquark models.
The presence of light right-handed neutrinos can affect these limits, a possibility that can occur for lepton number violation at a TeV, and that can be excluded with an improved bound
on $\mathcal{B}(D^0 \to \text{invisibles})$ at the level of $\sim 10^{-6}$, about two orders of magnitude better than the present one.
Signatures of $c \to u \nu \bar \nu$ modes contain missing energy and are suited for experimental searches at $e^+ e^-$--facilities, such as BES III, 
Belle II and future $e^+ e^-$--colliders, such as the FCC-ee running at the $Z$.
\end{abstract}

\maketitle

\section{Introduction}

Rare charm decays test physics beyond the standard model (BSM) and complement flavor studies with $K$'s and $B$'s in a unique way. 
An important tool in flavor and BSM searches are null tests -- observables that are very small in the standard model (SM)
due to approximate symmetries or parametric suppression.
Null tests allow to  bypass resonance backgrounds, which in the charm sector can be challenging otherwise~\cite{deBoer:2018buv}. Flavor changing neutral current (FCNC) charm dineutrino $c\to u \,\nu \bar \nu$ transitions are such null tests; being strongly Glashow-Iliopoulos-Maiani (GIM)-suppressed in the SM, their  branching ratios are tiny, such that any observation with current experimental sensitivities would cleanly signal new physics (NP)~\cite{Burdman:2001tf}.

\addtolength{\tabcolsep}{2pt} 
\begin{table}[h!]
 \centering
 \begin{tabular}{lccc}
  \hline
  \hline
  $h_c$  &  $f(c \to h_c)$  & $N(h_c)$  (a) & $N(h_c)$  (b) \\
  \hline
  $D^0$  &  $0.59$  &  $6 \cdot  10^{11}$ & $8 \cdot  10^{10}$   \\
  $D^+$  &  $0.24$  &  $3 \cdot  10^{11}$ & $3 \cdot  10^{10}$  \\
  $D_s^+$  &  $0.10$ &   $1 \cdot  10^{11}$ & $1 \cdot  10^{10}$   \\
  $\Lambda_c^+$  &  $0.06$ & $7 \cdot  10^{10}$ & $8 \cdot  10^{9}$\\
  \hline
  \hline
   \end{tabular}
   \caption{Charm fragmentation fractions $f(c \to h_c)$ \cite{Lisovyi:2015uqa} and  the number of charmed hadrons $h_c$, $N(h_c)$, expected at benchmarks
   with $N(c \bar c)=550 \cdot 10^{9}$ (a, FCC-ee) and $N(c \bar c)=65 \cdot 10^{9}$ (b, Belle II with $50\,\text{ab}^{-1}$) \cite{Abada:2019lih}. In absence of further information  for the $\Xi_c^+$  we use $f(c \to \Xi_c^+) \simeq f(c \to \Lambda_c^+)$. }
   \label{tab:frag}
\end{table}
\addtolength{\tabcolsep}{-2pt} 

The corresponding missing energy modes are well-suited for a clean $e^+ e^-$--collider environment, such as Belle II \cite{Kou:2018nap}, BES III \cite{Ablikim:2019hff}, and future colliders, notably, FCC-ee running at the $Z$ \cite{Abada:2019lih} with sizable charm production rates from $\mathcal{B}(Z \to c \,\bar c)\simeq 0.12$~\cite{Tanabashi:2018oca}. 
Fragmentation fractions $f(c \to h_c)$ of a charm quark to a charmed hadron $h_c$ from Ref.~\cite{Lisovyi:2015uqa} are compiled in TABLE~\ref{tab:frag}, together with the number of charmed hadrons $N(h_c)=2\,f(c\to h_c)\,N(c\bar{c})$ for FCC-ee and Belle II benchmark $c \bar c$ numbers \cite{Abada:2019lih}, (a) $N(c \bar c)=550 \cdot 10^{9}$ and (b) $N(c \bar c)=65 \cdot 10^{9}$, respectively.
With charmed hadron numbers of $\sim 10^{10}$ and higher, TABLE~\ref{tab:frag} reveals the potential of the $e^+ e^-$--machines for charm physics.

\begin{figure*}[!t]
\centering 
\includegraphics[width=0.95\textwidth]{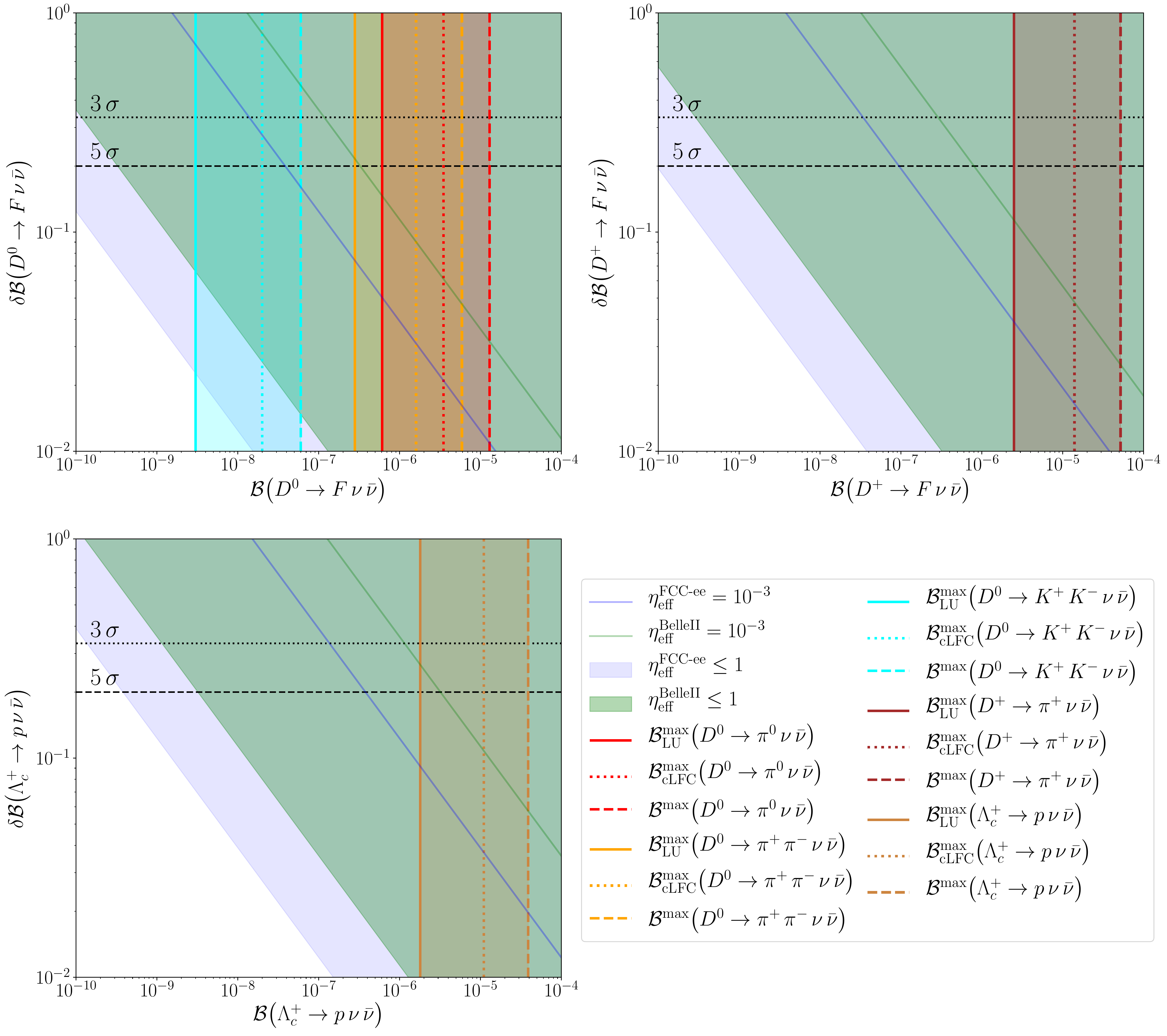}
\caption{Relative statistical uncertainty of the branching ratio $\delta\mathcal{B}$  versus the branching ratio  $\mathcal{B}$ for decays of 
the $D^0$ (upper plot to the left), the $D^+$ (upper plot to the right) and the $\Lambda_c^+$ (lower plot to the left). 
The shaded areas correspond to the reach for $\eta_\text{eff}=1$, whereas the solid tilted lines illustrate the impact of reconstruction efficiencies 
$\eta_\text{eff}=10^{-3}$ for the FCC-ee (lilac) and Belle II (green).
Horizontal $3 \,\sigma$  (dotted) and $5\,\sigma$  (dashed) black  lines correspond to $\delta\mathcal{B} =1/3$ and $\delta\mathcal{B} =1/5$, respectively.
Vertical lines represent upper limits assuming LU (solid), cLFC (dotted) and generic lepton flavor (dashed) for different modes, given in TABLE~\ref{tab:brbounds}. 
To improve readability the three lines for each decay mode are grouped together by a shaded band.
Upper limits for $D_s^+ \to K^+\,\nu\bar{\nu}$\,, $\Xi_c^+ \to \Sigma^+\,\nu\bar{\nu}$ and the inclusive modes can be seen in TABLE~\ref{tab:brbounds}.}\label{fig:relativebranchingDdecays}
\end{figure*}

To further detail the future sensitivities, 
we compute the expected event rate $N_{F}^{\text{exp}}$
for a decay $h_c \to F \nu \bar \nu$ with a final hadronic state $F$, as
\begin{align}
 N_{F}^{\text{exp}}\,=\,\eta_{\text{eff}}\,N(h_c)\,\mathcal{B}(h_c\to\, F\, \nu\,\bar\nu)\,, \label{eq:Num_hc}
\end{align}
where $\eta_{\text{eff}}$ accounts for the reconstruction efficiency. 
The relative statistical uncertainty for the branching ratio $\delta\mathcal{B}$ scales as $1/\sqrt{N_{F}^{\text{exp}}}$. 
In FIG.~\ref{fig:relativebranchingDdecays} we show the relative uncertainty $\delta\mathcal{B}$ against the branching ratio $\mathcal{B}$ for decays of the $D^0$ (upper plot to the left), the $D^+$ (upper plot to the right) and the $\Lambda_c^+$ (lower plot to the left). 
Since the fragmentation fractions of $\Lambda_c^+$ and $D_s^+$ are very similar the corresponding plot for $D_s^+$-mesons is not shown.
The left-most boundaries of the shaded regions correspond to the ideal, no-loss case $\eta_\text{eff}=1$, whereas the tilted lines illustrate the impact of reconstruction efficiencies of a permille
for the FCC-ee (lilac) and Belle II (green).
FIG.~\ref{fig:relativebranchingDdecays} demonstrates once more the high physics reach with sensitivities to (very) rare charm decays.
For efficiencies of a permille or better, branching ratios of  $\mathcal{O}(10^{-6})$ down to $\mathcal{O}(10^{-8})$ can be discovered  in $D^0$, $D^+_{(s)}$ and $\Lambda_c^+$ modes
at the (future) experiments, Belle II and FCC-ee. 
If sound estimates of $\eta_\text{eff}$ and systematic uncertainties would be available the reach could be determined in a more quantitative way. 
Here we stress that the region of branching ratios of $\mathcal{O}(10^{-6} -10^{-5})$ covers already interesting physics.
Note, since the displayed relation $\delta\mathcal{B}=1/\sqrt{ \eta_{\text{eff}}\,N(h_c)\,\mathcal{B}}$ does not depend on the final state, the estimated reach holds not only for dineutrino modes but also for radiative rare charm decays, with similar rates, {\it e.g.},  \cite{Burdman:2001tf,deBoer:2017que,Adolph:2020ema}.

Interestingly, there are no experimental upper limits on any of the $h_c \to F \nu \bar \nu$ modes available today.
However, recently, upper limits have been obtained using $SU(2)_L$-invariance and bounds on charged lepton modes~\cite{Bause:2020auq}.
In this work we provide further details on the implications of  this model-independent method.
Upper limits from the latter are shown as vertical lines in FIG.~\ref{fig:relativebranchingDdecays}.
For a given decay mode, the  upper limits depend on the charged lepton flavor structure: they are largest in the general case (dashed), followed by those assuming charged lepton flavor conservation (cLFC) (dotted) and if lepton universality (LU) holds (solid). To improve readability the three lines for each decay mode are grouped together by a shaded band.
The relevant ranges are suitable for Belle II and FCC-ee:  
all limits are above $\sim 10^{-6}$, with only one exception ($D^0 \to K^+ K^- \nu \bar \nu $).
Upper limits for $D_s^+ \to K^+\,\nu\bar{\nu}$\,, $\Xi_c^+ \to \Sigma^+\,\nu\bar{\nu}$ and the inclusive modes are provided in TABLE~\ref{tab:brbounds}.
The hierarchy of upper limits per mode allows for the exciting possibility to probe charged lepton flavor properties using fully flavor-summed dineutrino branching ratios with unreconstructed neutrino flavor. 
This concludes our introduction and motivation to work  out  the physics reach of charmed dineutrino modes.

The plan of the paper is as follows:
We introduce the weak effective hamiltonian for $c \to u \nu \bar \nu$ transitions in Section \ref{sec:efftheofram}.
In Section \ref{sec:branchingratios} we analyze the decay distributions of $D_{(s)} \to P \,\nu \bar \nu$, $D _{(s)} \to P^+ P^-  \,\nu \bar \nu$, $P=\pi, K$ $\Lambda_c^+ \to p\, \nu \bar \nu$, $\Xi_c^+ \to \Sigma^+\,  \nu \bar \nu$  and inclusive modes $D \to X  \nu \bar \nu$.
We obtain model-independent predictions for branching ratios in Section \ref{sec:modelindependbounds}.
We also consider  the implications and constraints from right-handed (RH)  neutrinos and lepton number violating (LNV) interactions in the charm sector.
Predictions for tree-level NP mediators, such as $Z^\prime$ and leptoquark (LQ) models are discussed in Section \ref{sec:tree-level}.
We conclude in Section \ref{sec:con}.  
Appendix \ref{sec:detailsu2} provides details on the $SU(2)_L$-link and probing LU and cLFC. Appendix \ref{app:FF} contains formulae for form factors.

\section{Low-energy effective Hamiltonian}\label{sec:efftheofram}

In the absence of light RH neutrinos, as in the SM, $|\Delta c|=|\Delta u|=1$ dineutrino transitions can be described by two operators amended by flavor indices in the weak effective hamiltonian
\begin{equation} \label{eq:Heff_noright}
	\mathcal H_\text{eff}^{\nu_i  \bar \nu_j} = -\frac{4\,G_\text F}{\sqrt 2}\frac{\alpha_e}{4\pi} \left( \mathcal{C}_L^{Uij} Q_L^{ij} + \mathcal{ C}_R^{Uij}Q_R^{ij} \right) + \,\text{H.c.}\,,
\end{equation}
with the four-fermion operators
\begin{align}\label{eq:opsneu} 
Q_{L(R)}^{ij}&=  (\bar u_{L(R)} \gamma_\mu c_{L(R)})  \,  (\bar \nu_{jL} \gamma^\mu \nu_{iL}) \, , 
\end{align}
and $i,j$ denote the neutrino flavors (mass eigenstates).  Here, $G_F$ denotes Fermi's constant and $\alpha_e$ is the fine structure constant.
No further dimension six operators exist in $\mathcal H_\text{eff}^{\nu_i  \bar \nu_j}$.

Since the neutrino flavor indices are not experimentally tagged, dineutrino branching ratios are obtained by adding all dineutrino flavors incoherently
\begin{align} \label{eq:sum}
\mathcal{B}( c \to u \,\nu \bar \nu)=\sum_{i,j} \mathcal{B}( c \to u \,\nu_j \bar \nu_i) \, . 
\end{align}
Therefore, all branching ratios depend on at most two combinations of Wilson coefficients that can be chosen as
\begin{align}
    x_U^\pm=  \sum_{i,j}  \big\vert\mathcal{C}_L^{U ij}\pm\mathcal{C}_R^{U ij}\big\vert^2  \, . 
\end{align}
As it enters inclusive rates, the following term turns out to be useful for the discussion of experimental limits
\begin{align}\label{eq:xz}
    x_U = \frac{x_U^+ + x_U^-}{2}= \sum_{i,j}\left(\vert\mathcal{C}_L^{U ij}\big\vert^2+\vert\mathcal{C}_R^{U ij}\big\vert^2\right)\,.
\end{align}
$x_U$, and therefore $x_U^\pm \leq 2\, x_U$, are presently not constrained by direct experimental information on $|\Delta c|=|\Delta u|=1$ dineutrino transitions.
On the other hand, model-independent upper limits on $x_U$ have been
derived using $SU(2)_L$-invariance and data on charged lepton processes~\cite{Bause:2020auq}.
With upper limits depending on the charged lepton flavor structure, limits are obtained in three scenarios:
LU, cLFC and general  lepton flavor structure.

Specifically, writing the weak effective hamiltonian for charged dileptons as
\begin{equation}\label{eq:Heff_opsCH}
	\mathcal{H}_\text{eff}^{\ell_i \ell_j} \supset -\frac{4\,G_{\text{F}}}{\sqrt 2}\frac{\alpha_e}{4\pi} \left( \mathcal{K}_L^{U ij} O_L^{ij} + \mathcal{K}_R^{U ij}O_R^{ij}\right) +\text{H.c.},
\end{equation}
with dileptonic operators
\begin{align}\label{eq:opsCH}
\begin{split}
    	O_{L(R)}^{ij}&=  (\bar u_{L(R)} \gamma_\mu c_{L(R)} ) \, (\bar \ell_{jL} \gamma^\mu \ell_{iL}) \, , 
\end{split}
\end{align}
analogous to the weak hamiltonian for dineutrinos \eqref{eq:Heff_noright},  the LU, cLFC limits corresponding to flavor structures in the Wilson coefficients 
can be identified as
\begin{align}  \label{eq:patterns}
\mathcal{K}_{L,R}^{U}|_\text{LU}\! = \! \begin{pmatrix} k  &0 & 0\\ 0 & k &0 \\ 0 & 0 & k \end{pmatrix} ,~ 
\mathcal{K}_{L,R}^{U}|_\text{cLFC} \!= \! \begin{pmatrix} k_e  &0 & 0\\ 0 & k_\mu &0 \\ 0 & 0 & k_\tau \end{pmatrix}, 
\end{align}
while ``general'' means that all entries in the coefficient matrix are arbitrarily filled, allowing for cLFV.
Here, $k, k_\ell$ are the parameters in the coefficient matrix. 

Given a relation~\cite{Bause:2020auq} between the neutrino $\mathcal{C}_{L,R}^{ij}$ and the charged lepton $\mathcal{K}_{L,R}^{ij}$ couplings bounds on the latter imply limits on the former. 
Note, this relation involves also down-sector couplings to charged leptons,
$\mathcal{K}_{L,R}^{D}$, with analogous flavor patterns as in the up-sector \eqref{eq:patterns}. 
Clearly the limits depend on the flavor structure.

Using input provided in Appendix \ref{sec:detailsu2}, to which we also refer for details, the  upper limits read
\begin{align}  \label{eq:LU1}
x_U &
\lesssim 34\,, \,\,\quad  (\text{LU}) \\ \label{eq:cLFC1}
x_U &
 \lesssim 196 \,, \quad (\text{cLFC}) \\ \label{eq:total1}
x_U &
\lesssim  716  \,, \quad (\text{general}) \,,   
\end{align}
which include leading order corrections from the Wolfenstein parameter $\lambda\simeq 0.2$, therefore providing larger upper limits than in Ref.~\cite{Bause:2020auq}.
We employ these model-independent, data-driven limits in the following Section \ref{sec:branchingratios} as benchmarks for differential decay distributions.
In Section \ref{sec:modelindependbounds} we present upper limits on the branching ratios using \eqref{eq:LU1}-\eqref{eq:total1}. 
We also discuss the impact of RH neutrinos. 

We remark that the charged lepton data yielding \eqref{eq:LU1}-\eqref{eq:total1} are from LHC's Drell-Yan studies~\cite{Fuentes-Martin:2020lea,Angelescu:2020uug}.
In contrast to constraints from rare decays, here operators do not interfere and large cancellations are avoided. 
On the other hand,  especially  in the
down-sector rare decay data can imply significantly stronger constraints. Yet, as  discussed in Appendix \ref{sec:detailsu2},  the  upper limits  on the $x_U$ 
including kaon constraints remain  within the same
order of magnitude  as in  \eqref{eq:LU1}-\eqref{eq:total1}. Therefore, 
we choose  total model-independence and  conservatively present results for   \eqref{eq:LU1}-\eqref{eq:total1}.

\section{Differential branching ratios}\label{sec:branchingratios}

The differential branching ratios of the dineutrino modes can be written as
\begin{align}\label{eq:dBR}
    \frac{\text{d}\,\mathcal{B}(h_c\to F\,\nu \bar \nu)}{\text{d} \,q^2} =\phantom{+}\, a_+^{h_c F}(q^2)\,x_U^+ +\,a_-^{h_c F}(q^2)\,x_U^-\,,
\end{align}
where $q^2$ denotes the invariant mass-squared of the dineutrinos. Eq.~\eqref{eq:dBR} can also be expressed in terms of missing energy, that is the energy of neutrinos, in the charmed hadron's center-of-mass system, as $\text{d} \mathcal{B}/ \text{d} E_\text{miss}=2\, m_{h_c} \text{d} \mathcal{B}/\text{d} q^2$, where $m_{h_c}$ denotes the mass of the initial charm hadron. 
The $q^2$--dependent functions $a_\pm^{h_c F}$ can be fetched from the literature \cite{Bause:2019vpr,deBoer:2018buv,Das:2018sms,Altmannshofer:2009ma,Boer:2014kda}, and are given in Sections \ref{sec:DP}-\ref{sec:incl}. 
Information on the form factors from Refs.~\cite{Lubicz:2017syv,Lee:1992ih,deBoer:2018buv,Meinel:2017ggx} is compiled  in Appendix~\ref{app:FF}.

\begin{table}[ht!]
 \centering
 \begin{tabular}{lcccccccccc}
  \hline
  \hline
  $h_c\to F $ & $A^{h_c\,F}_+$ & $A^{h_c\,F}_-$  \\
  &$[10^{-8}]$&$[10^{-8}]$\\
  \hline
  $D^0\to\pi^0$ & 0.9 & 0  \\
  $D^+\to\pi^+$ & $3.6$ & 0\\
  $D_s^+\to K^+$ & $0.7$ & 0 \\
  && \\
  $D^0\to\pi^0\pi^0$ & $0.7 \cdot 10^{-3}$ & $0.21$  \\
  $D^0\to\pi^+\pi^-$ & $1.4 \cdot 10^{-3}$ & $0.41$ \\
  $D^0\to K^+K^-$ & $4.7 \cdot 10^{-6}$ & 0.004\\
  && \\
  $\Lambda_c^+\to p^+$ & 1.0 & 1.7 \\
  $\Xi_c^+\to \Sigma^+$ & 1.8 & 3.5  \\ 
  && \\
  $D^0\to X $ & $2.2$ & $2.2$  \\
  $D^+\to X $ & $5.6$ & $5.6$ \\
  $D_s^+\to X$ & $2.7$ & $2.7$ \\
  \hline
  \hline
  \end{tabular}
  \caption{Coefficients $A^{h_c\,F}_\pm$ as in Eq.~\eqref{eq:tapm} for various  charmed hadrons  $h_c$ and final states $F$ for central values of
  input.
  For the exclusive charged $D$ decays $q^2$-cuts \eqref{eq:cut} are taken into account, while for inclusive modes
  no cuts were applied, as the details of possible backgrounds are beyond the scope of this work. This table is adopted from Ref.~\cite{Bause:2020auq}.
  }
  \label{tab:afactors}
\end{table}

Integrating the differential branching ratios Eq.~\eqref{eq:dBR}, one finds
\begin{align}\label{eq:BR}
    \mathcal{B}(h_c\to F\,\nu \bar \nu)= A_+^{h_c F}\,x_U^+\,+ A_-^{h_c F}\,x_U^-\,,
\end{align}
where 
\begin{align}\label{eq:tapm}
    A_\pm^{h_c F}&=\int_{q^2_{\text{min}}}^{q^2_{\text{max}}}\,\text{d}q^2\,a_\pm^{h_c F}(q^2)\,.
\end{align}
Here, $q^2_{\text{max}}=(m_{h_c}-m_F)^2$ for the exclusive modes and
$q^2_{\text{max}}=m_c^2$ for inclusive $D^{0,+}$ and
$q^2_{\text{max}}=(m_D-m_K)^2$  for inclusive $D_s^+$ decays~\cite{Buchalla:1998mt}.
$m_F$ ($m_D$) denotes the mass of the hadronic final state ($D$-meson). 
For two pseudoscalars $F=P_1 P_2$, 
$P_{1,2}=\pi,K$, $m_F=m_{P_1}+m_{P_2}$, where $m_{P_i}$ denotes the mass of the pseudoscalar meson $P_i$.
Resonant backgrounds in charged meson decays through $\tau$--leptons, {\it i.e.,} $D^+ \to \tau^+ (\to \pi^+ \bar \nu) \nu$ and $D_s^+ \to \tau^+ (\to K^+ \bar \nu) \nu$ lead to the same final state as the search channels $D^+ \to  \pi^+ \bar \nu \nu$ and $D_s^+ \to  K^+ \bar \nu \nu$~\cite{Kamenik:2009kc}
and need to be removed by kinematic cuts 
\begin{equation}
     q^2>(m_\tau^2-m_P^2)(m_D^2-m_\tau^2)/m_\tau^2\,,\label{eq:cut}
\end{equation}
where $m_\tau$ denotes the mass of the tau.
Therefore, the integration region in \eqref{eq:tapm} is bounded by $q^2_{\text{min}}= 0.34\,\text{GeV}^2 \,(0.66\,\text{GeV}^2)$ for $D^+\to\pi^+ \bar \nu\nu$ ($D^+_s\to K^+ \bar \nu \nu$), 
whereas we use $q^2_{\text{min}}=0$ in all other modes. 
We note that the inclusive decays require phase space cuts, however, a dedicated analysis of an experimental strategy is beyond the scope of this work.

In TABLE~\ref{tab:afactors} we provide the central values for the prefactors $A_\pm^{h_c F}$, taking into account Eq.~\eqref{eq:cut} for exclusive $D^+$ and $D_s^+$-decays.
As expected from Lorentz-invariance and parity-conservation in the strong interaction we observe
\begin{enumerate}[label=(\alph*)]
    \item $A_-^{h_c F}= 0$ in $D\to P\, \nu \bar \nu$ decays,
    \item $A_+^{h_c F}\ll  A_-^{h_c F}$ in $D\to P_1 P_2\, \nu \bar \nu$ decays,
    \item $\mathcal{O}(A_-^{h_c F})\sim\mathcal{O}(A_+^{h_c F})$ in baryonic charm decays,
    \item $A_-^{h_c F} = A_+^{h_c F}$ in inclusive $D$ decays,
\end{enumerate}
which highlights the complementarity between the different decay modes in regard of NP sensitivity. 
We return to this in Section \ref{sec:corr}.

In the following sections \ref{sec:DP}-\ref{sec:incl} we review the theory description of the decays $D\to P\, \nu\bar\nu$, $D\to P_1 P_2 \,\nu\bar\nu$,
$\Lambda_c^+\to p\,\nu\,\bar \nu$ and $\Xi_c^+ \to \Sigma^+ \, \nu \bar \nu$ and $D\to X \, \nu\bar\nu$
and provide details relevant for the calculation of the $A^{h_cF}_\pm$ factors compiled in TABLE~\ref{tab:afactors}.

\subsection{$\boldsymbol{D\to P\, \nu\bar\nu}$ \label{sec:DP}}

The $D\to P \,\nu \bar\nu$ mode, where $D=D^0,\,D^+,\,D^+_s$ and $P=\pi^0,\,\pi^+,\,K^+$, respectively, is described by only one form factor.
The  $a_\pm^{D P}$--functions of the differential decay width can be written as
\begin{align}
a_+^{D P}(q^2) = \frac{G_{\text{F}}^2\, \alpha_e^2\,\tau_D {\lambda(m_D^2,m_P^2,q^2)}^{\frac{3}{2}}\, ({f_+^{DP}(q^2)})^2}{3072\,\pi^5\, m_D^3}\,,
\label{eq:dGamma}
\end{align}
and $a_-^{D P}(q^2) =0$.
Here, $\lambda(a,\,b,\,c)=a^2+b^2+c^2-2 \,(ab-ac-bc)$ is the usual Källén function
and $\tau_D$ denotes the lifetime of the $D$-meson.

In this work we use the $D\to P$ form factors computed  by Lubicz et al.~\cite{Lubicz:2017syv} using lattice QCD. 
Details can be found in Appendix~\ref{app:ff_DP}. 
FIG.~\ref{fig:dBRDPnunu} illustrates the differential branching ratio for all three decay modes with exemplary values of $x_U$ from Eqs.~\eqref{eq:LU1} (solid) and \eqref{eq:cLFC1} (dotted).

\begin{figure}[h!]
\centering
\includegraphics[width=0.45\textwidth]{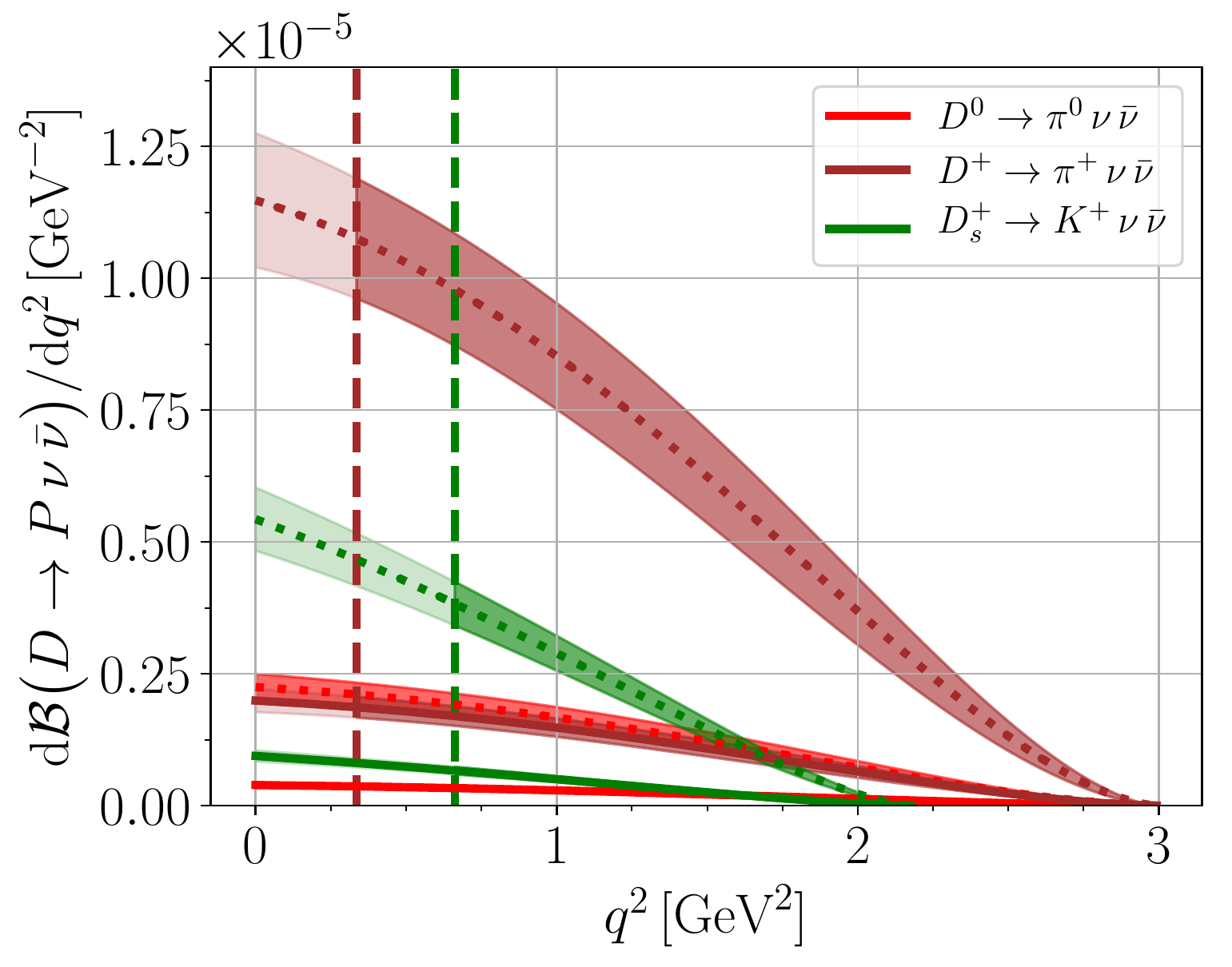}
\caption{Differential branching ratios for $D^0\to \pi^0 \nu\bar\nu$, $D^+\to \pi^+ \nu\bar\nu$ and $D_s^+\to K^+ \nu\bar\nu$ in red, brown and green, respectively for the LU (cLFC) limit in solid (dotted) lines from Eq.~\eqref{eq:LU1} (Eq.~\eqref{eq:cLFC1}). 
The uncertainty bands are due to the form factors, the vertical dashed lines illustrate the cuts \eqref{eq:cut} needed to avoid the $\tau$ background.}
\label{fig:dBRDPnunu}
\end{figure}

The hierarchy between the $D^0,\,D^+,\,D^+_s$
branching ratios stems predominantly from the different lifetimes
$\tau_{D^0} < \tau_{D_s} < \tau_{D^+}$, while for the $D_s$ also the phase space difference relative to the $D^{0,+}$ plays a role.

\subsection{$\boldsymbol{D\to P_1 P_2 \,\nu\bar\nu}$}

The angular distributions of $D\to P_1\, P_2\, \nu\bar \nu$ decays can be obtained from Ref.~\cite{deBoer:2018buv}. 
Integrating the unobservable kinematic variables,
two angles in the full five-fold angular distribution, 
we are left with a three-fold differential distribution with the following $a_\pm^{D P_1 P_2}$--functions,
\begin{align}\label{eq:apmDPP}
    a_\pm^{D P_1 P_2}(q^2)=\int_{(m_{P_1}+m_{P_2})^2}^{(m_D-\sqrt{q^2})^2}\hspace{-0.1cm}\text{d}p^2\hspace{-0.1cm}\int_{-1}^{1}\hspace{-0.2cm}\text{d}\cos\theta_{P_1} b_\pm(q^2,p^2,\theta_{P_1})\,,
\end{align}
with 
\begin{align}\label{eq:I1I2} 
\begin{split}
b_-(q^2,p^2,\theta_{P_1}) & = \frac{\tau_D}{6}\,\bigg[|\mathcal{F}_0 |^2 + \sin^2 \theta_{P_1} |\mathcal{F}_\parallel |^2\bigg]  \,,\\ 
b_+(q^2,p^2,\theta_{P_1}) & = \frac{\tau_D}{6}\,\sin^2 \theta_{P_1} |\mathcal{ F}_\perp |^2 \,,
\end{split}
\end{align}
where $p^2$ denotes the invariant mass-squared of the ($P_1 P_2$)-subsystem.
$\theta_{P_1} $ is the angle between the $P_1$-momentum and the negative direction of flight of the $D$-meson in the ($P_1P_2$)-cms. 
The transversity form factors $\mathcal{F}_i$, $i=0, \perp, \parallel$ are given in Ref.~\cite{deBoer:2018buv},  with details provided
 in Appendix~\ref{app:ff_DP1P2}.

Helicity relations imply that at low hadronic recoil the transverse perpendicular form factor is suppressed with respect to the others, 
$\mathcal{F}_\perp \ll \mathcal{F}_{0, \parallel}$ \cite{Hiller:2013cza}. In addition, at large recoil the longitudinal form factor becomes the leading one, 
$\mathcal{F}_{\perp , \parallel}  \ll \mathcal{F}_{0}$. Therefore, $D\to P_1\, P_2\, \nu\bar \nu$ decays are 
dominated by the  $A_-^{D P_1 P_2}$ contribution, as can be seen numerically in TABLE~\ref{tab:afactors}, and have only suppressed sensitivity to  $x_U^+$.

\begin{figure}[h!]
    \centering
\includegraphics[width=0.45\textwidth]{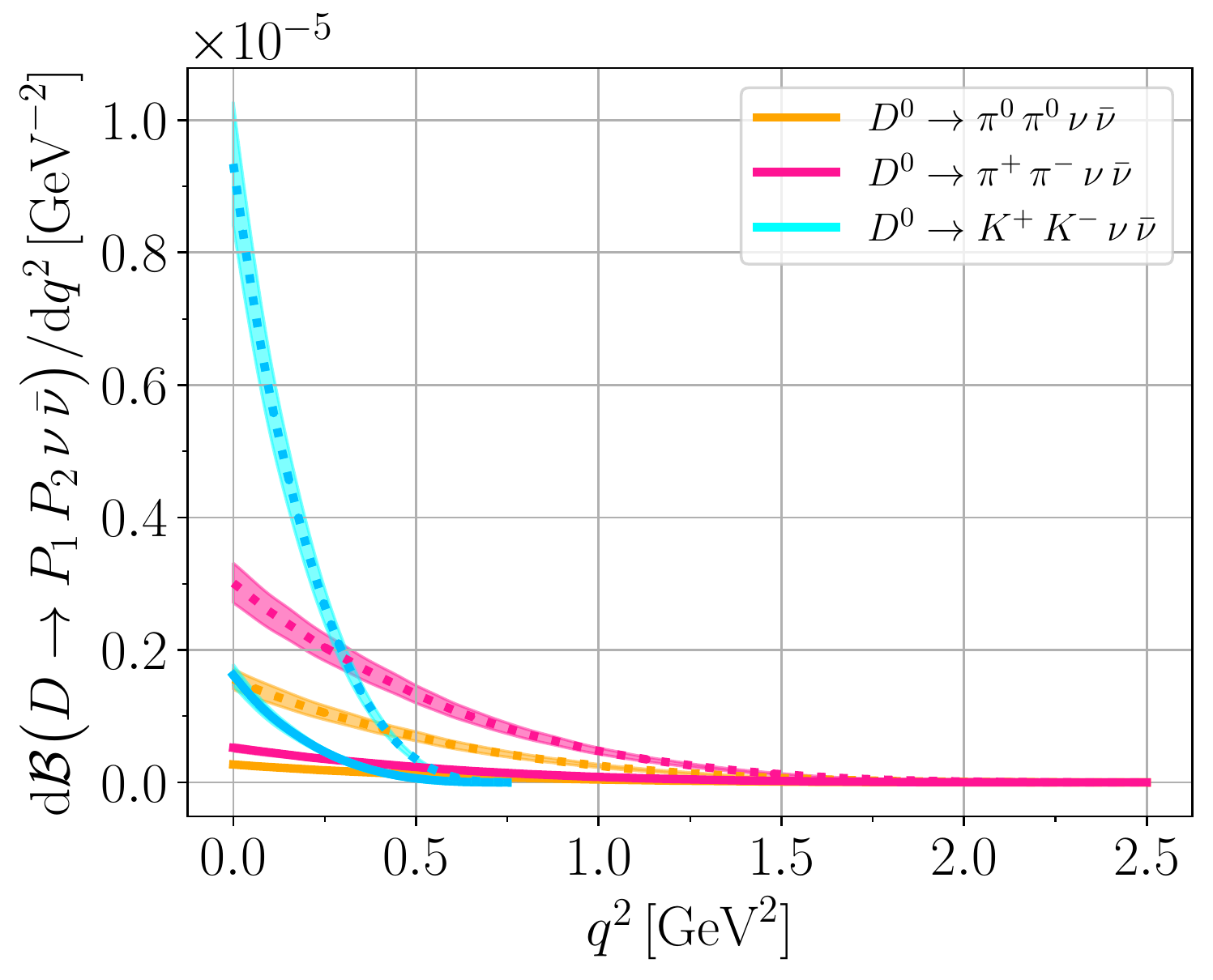}
\includegraphics[width=0.45\textwidth]{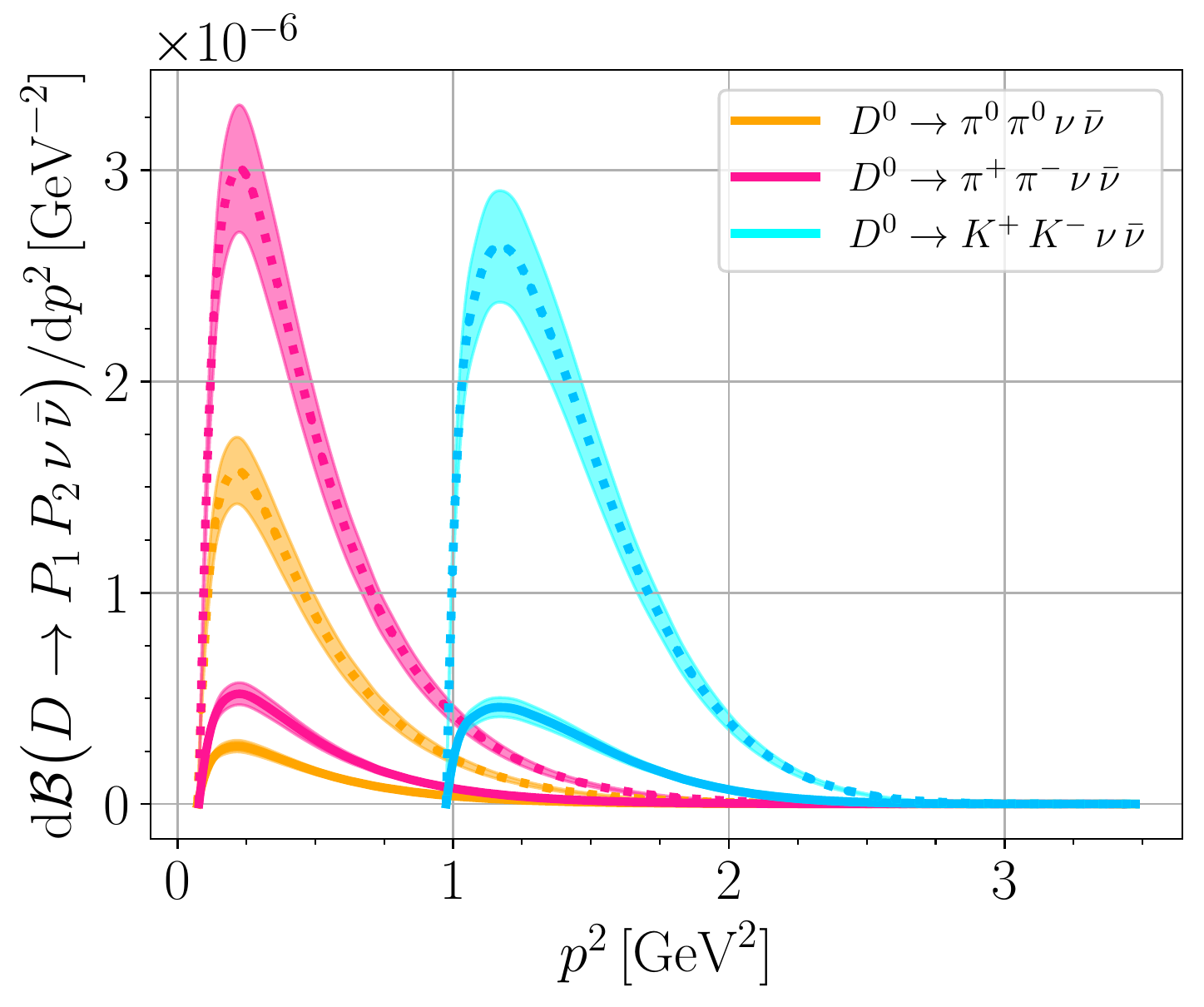}
\caption{Differential branching ratios for $D^0\to\pi^0\,\pi^0\,\nu\bar\nu$, $D^0\to\pi^+\,\pi^-\,\nu\bar\nu$ and $D^0\to K^+\,K^-\,\nu\bar\nu$ decays in orange, deep pink and cyan, respectively for the LU (cLFC) limit in solid (dotted) lines from Eq.~\eqref{eq:LU1} (Eq.~\eqref{eq:cLFC1}). 
The upper plot shows $\text{d}\mathcal{B}/{\text{d} q^2}$, whereas the lower plot $\text{d}\mathcal{B}/{\text{d} p^2}$, as in \eqref{eq:explainp2}. 
The differential branching ratio of $D^0\to K^+\,K^-\,\nu\bar\nu$ is multiplied by a factor 100 to be visible in the plots. 
The band widths illustrate $10\,\%$ uncertainty originating from form factors.}
\label{fig:dBRDP1P2}
\end{figure}

In FIG.~\ref{fig:dBRDP1P2} (upper plot) we illustrate the $q^2$-differential branching ratio for three decay modes, $D^0\to\pi^0\,\pi^0\,\nu\bar\nu$, $D^0\to\pi^+\,\pi^-\,\nu\bar\nu$ and $D^0\to K^+\,K^-\,\nu\bar\nu$, for  $x_U$ saturating  Eqs.~\eqref{eq:LU1} and \eqref{eq:cLFC1}.
Also shown  are the ($P_1-P_2$)-mass-squared distributions  $d\mathcal{B}/{d p^2}$, obtained as
\begin{align}
\label{eq:explainp2}
  \frac{\text{d}\,\mathcal{B}(D\to P_1 P_2 \,\nu \bar \nu)}{\text{d} \,p^2} = \phantom{+}\, a_+^{D P_1P_2}(p^2)\,x_U^+&  +\,a_-^{D P_1P_2 }(p^2)\,x_U^-\,, 
  \end{align}
  \begin{align}
    a_\pm^{D P_1 P_2}(p^2)=  \int_{0}^{(m_D-\sqrt{p^2})^2}\hspace{-0.3cm}\text{d}q^2 \hspace{-0.1cm}\int_{-1}^{1} & \hspace{-0.2cm}\text{d}\cos\theta_{P_1} b_\pm(q^2,p^2,\theta_{P_1})\,,
\end{align}
in close analogy to  \eqref{eq:dBR} and \eqref{eq:apmDPP}.
Due to isospin, the distributions for $F=\pi^0 \pi^0$ and $\pi^+ \pi^-$ are essentially the same up to an overall factor of 2, due to two identical particles in the final state.

\subsection{Charmed baryon modes \label{sec:baryons}}

\begin{figure}[h!]
\centering
\includegraphics[width=0.45\textwidth]{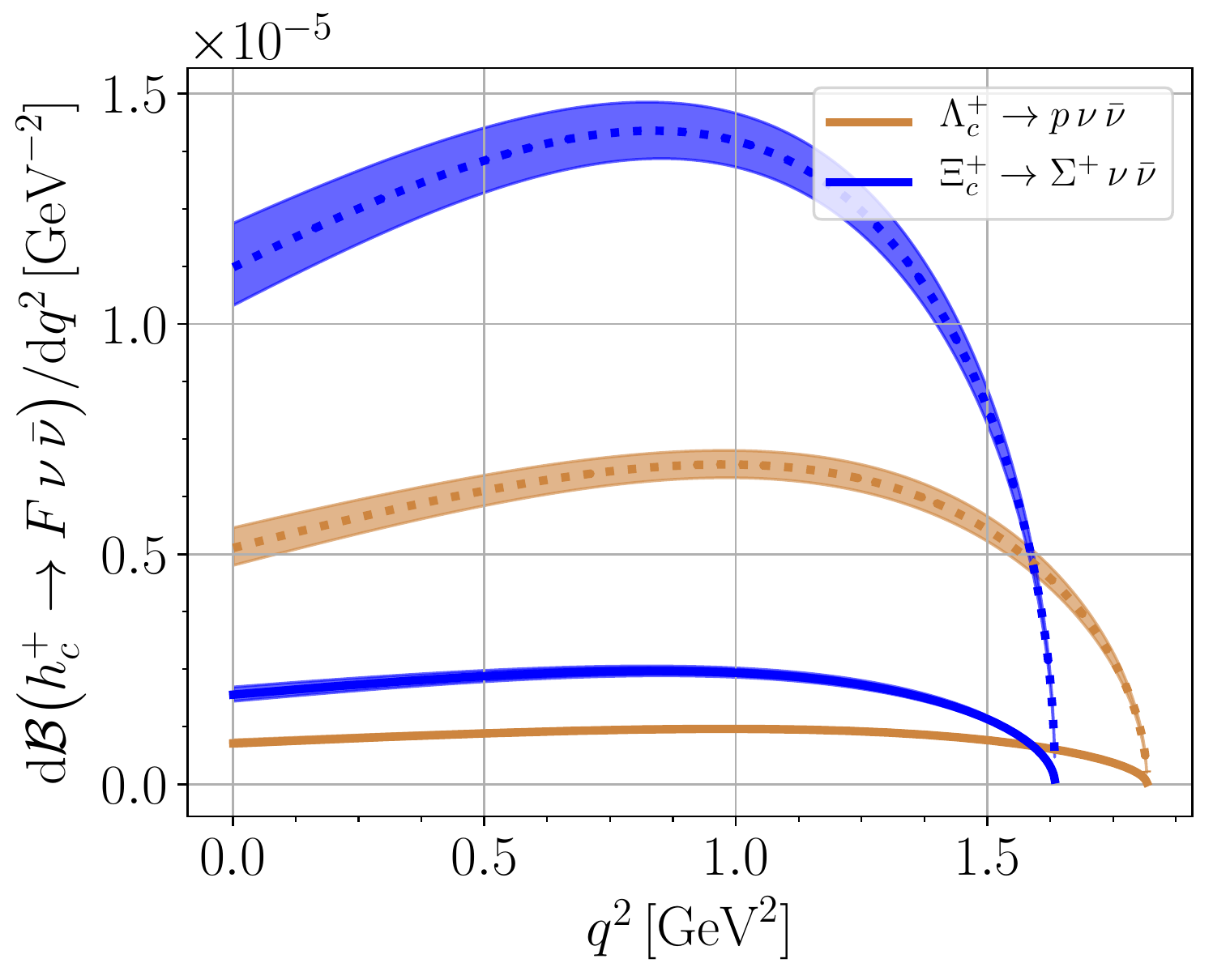}
\caption{Differential branching ratios for $\Lambda_c^+\to p\,\nu\,\bar \nu$ and $\Xi_c^+ \to \Sigma^+ \, \nu \bar \nu$ decays  in brown and blue, respectively for the LU (cLFC) limit in solid (dotted) lines from Eq.~\eqref{eq:LU1} (Eq.~\eqref{eq:cLFC1}). 
The band widths correspond to the form factor uncertainties, see main text.}
\label{fig:dBRlambda}
\end{figure}

The differential decay rates for $\Lambda_c^+\to p\,\nu\,\bar \nu$ and $\Xi_c^+ \to \Sigma^+ \, \nu \bar \nu$ decays can be extracted from Ref.~\cite{Das:2018sms}. 
Four different form factors enter the $a_\pm^{h_c^+ F}$--functions as 
\begin{align}
\begin{split}\label{eq:LamddacPM}
    a_+^{h_c^+F}(q^2)&=\mathcal{N}\bigg(2 f_{\perp}^2 s_-+f_+^2(m_{h_c^+}+m_F)^2  \frac{s_-}{q^2} \bigg)\,,\\
    a_-^{h_c^+ F}(q^2)&=\mathcal{N}\bigg(2 g_{\perp}^2 s_+ +g_+^2(m_{h_c^+}-m_F)^2  \frac{s_+}{q^2} \bigg)\,,
\end{split}
\end{align}
with $s_\pm = (m_{h_c^+} \pm m_F)^2 - q^2$ and
\begin{align}
\mathcal{N}=\frac{G_{\text{F}}^2\, \alpha_e^2\, \tau_{h_c^+}\,q^2\sqrt{\lambda(m_{h_c^+}^2,m_F^2,q^2)}}{2^{10}\,3\,m^3_{h_c^+} \pi^5 }  \,.
\end{align}
Here, $\tau_{h_c^+}$ ($m_{h_c^+}$) denote the lifetime (mass) of the charm hadrons.
For the charmed baryon modes, we use the form factors provided in Ref.~\cite{Meinel:2017ggx}. 
Details can be found in Appendix~\ref{app:ff_lambdac}. 
In view of missing computations for the $\Xi_c^+$ mode, we adopt the same form factors as for the $\Lambda_c^+$ one. 
FIG.~\ref{fig:dBRlambda} illustrates the differential branching ratio for these two decay modes for  $x_U$ saturating  Eqs.~\eqref{eq:LU1} and \eqref{eq:cLFC1}.
Within our working assumption of similar form factors, the decays of the $\Xi_c^+$ (blue) are  about twice as often than the $\Lambda_c^+$ ones (brown) due to the difference in lifetime, $\tau_{\Xi_c}/\tau_{\Lambda_c^+} \simeq 2$~\cite{Zyla:2020zbs}.

\subsection{Inclusive $\boldsymbol{D\to X\, \nu\bar\nu}$ decays}\label{sec:incl}

The $D\to X\, \nu\,\bar \nu$  decays with an inclusive hadronic final state with flavor quantum number of an up-quark, $X=\pi, \pi \pi, \ldots$\,, for $D^{0,+}$ decays
or an anti-strange quark from  $D_s^+$ decays,  $X=K , K  \pi, \ldots$\,, are complementary to the exclusive ones in several aspects: 
the theory framework for inclusive modes is an operator product expansion, rather than one involving form factors, and  in the different experimental analysis. 
In addition, inclusive modes are proportional to $x_U$. 
The corresponding dineutrino mass distribution can be written in terms of $a_\pm^{D X}$ as~\cite{Altmannshofer:2009ma}
\begin{align}\label{eq:DXunn}
    a_\pm^{D X}(q^2)=\frac{G_{\text{F}}^2 \,\alpha_e^2\,\tau_D \,m_c^3}{2^{10}\,3\, \pi^5}\, \kappa(0)\,f_{\text{incl.}}(q^2)\,,
\end{align}
where
\begin{align}
    f_{\text{incl.}}(q^2)\,=\,\bigg(1-\frac{q^2}{m_c^2}\bigg)^2\left[1+2\,\frac{q^2}{m_c^2}\right]\,,
\end{align}
and
\begin{align}
    \kappa(0)=1\,+\,\frac{\alpha_s(m_c)}{\pi}\left[\frac{25}{6}-\frac{2}{3}\,\pi^2\right]\approx0.71\,,
\end{align}
represents the QCD correction to the $c\to u\,\nu\bar\nu$ matrix element inferred from Ref.~\cite{Bobeth:2001jm}. 

\begin{figure}[h!]
\centering
\includegraphics[width=0.45\textwidth]{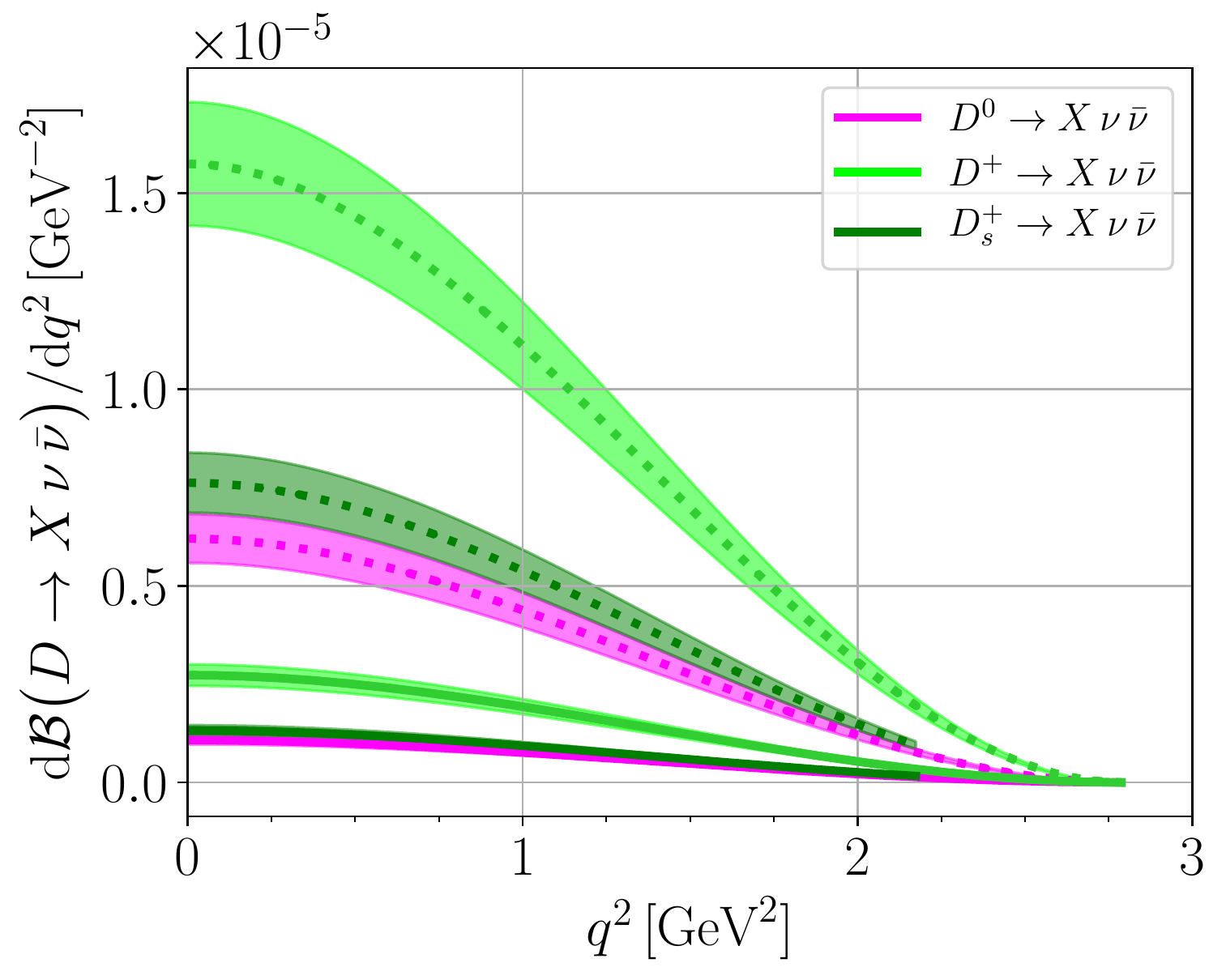}
\caption{Differential branching ratios for $D^0 \to X \nu \bar \nu$, $D^+ \to X  \nu \bar \nu$ and $D^+_s \to X  \nu \bar \nu$ decays in magenta, lime and green, respectively for the LU (cLFC) limit in solid (dotted) lines from Eq.~\eqref{eq:LU1} (Eq.~\eqref{eq:cLFC1}). 
The band widths illustrate $10\,\%$ uncertainties from power corrections. 
The distributions are cut at $q^2_\text{max}=m_c^2$ and at the physical limit $q^2_\text{max}=\left(m_D-m_K\right)^2$ for the $D^{0,+}$ and $D_s^+$ modes, respectively~\cite{Buchalla:1998mt}.
}
\label{fig:dBRDXu}
\end{figure}

FIG.~\ref{fig:dBRDXu} illustrates the differential branching ratio for three decay modes, $D^0 \to X  \nu \bar \nu$, $D^+ \to X  \nu \bar \nu$ and $D^+_s \to X  \nu \bar \nu$, for $x_U$ saturating  Eqs.~\eqref{eq:LU1} and \eqref{eq:cLFC1}.
We observe similar hierarchies between the decay modes as in FIG.~\ref{fig:dBRDPnunu}, which are driven by the lifetimes.

\section{Model independent tests}\label{sec:modelindependbounds}

We discuss model-independent tests of NP, their implications for flavor physics and potential challenges due to the presence of light BSM neutrinos.
Using the model-independent upper limits on the $|\Delta c|=|\Delta u|=1$ dineutrino Wilson coefficients in the  flavor benchmarks LU, cLFC and general 
flavor \eqref{eq:LU1}-\eqref{eq:total1}, together with the description of $h_c \to F \nu \bar \nu$ decays from Section \ref{sec:branchingratios},
we are now in the position to predict upper limits on branching ratios.
These are presented in Section \ref{sec:bounds}, together with implications for flavor and tests at $e^+ e^-$--machines.
We discuss the correlation between different decays in Section \ref{sec:corr}, that arises in an overconstrained system with more observables (decay modes) than unknowns ($x_U^\pm$). 
In Section \ref{sec:RH} we go beyond the assumption of SM-like light neutrinos and allow for right-handed neutrinos.
We discuss implications and constraints. 
RH-neutrinos can appear in models with low-scale see-saw origin of neutrino mass.
We work out constraints on LNV in Section \ref{sec:lnv}.

\subsection{Branching ratios probe NP and flavor \label{sec:bounds}}

Using the bounds on $x_U$ from Eqs.~\eqref{eq:LU1}, \eqref{eq:cLFC1} and \eqref{eq:total1}, together with Eq.~\eqref{eq:BR} and the values of $A_\pm^{h_c\,F}$ provided in TABLE~\ref{tab:afactors}, 
we obtain upper limits on branching ratios for the three flavor scenarios, that is, LU ($\mathcal{B}_{\text{LU}}^{\text{max}}$), cLFC ($\mathcal{B}_{\text{cLFC}}^{\text{max}}$) and general ($\mathcal{B}^{\text{max}}$). 
The maximal branching ratios are given in TABLE~\ref{tab:brbounds}
and have been shown in FIG.~\ref{fig:relativebranchingDdecays} for $D^0,D^+$ and $\Lambda_c^+$-decays. 
As already stressed in the Introduction, upper limits are in the right ballpark for study at Belle II and FCC-ee.
The upper limits satisfy  $\mathcal{B}_{\text{LU}}^{\text{max}}<\mathcal{B}_{\text{cLFC}}^{\text{max}}<\mathcal{B}^{\text{max}}$ and we recall that they
correspond to a specific flavor structure in the charged lepton sector. 
Then, for instance, a branching ratio measurement $\mathcal{B}_\text{exp}$ in some mode within $\mathcal{B}_{\text{LU}}^{\text{max}}<\mathcal{B}_\text{exp}<\mathcal{B}_{\text{cLFC}}^{\text{max}}$ would be a signal of LU violation. 
In contrast, a branching ratio above $\mathcal{B}_{\text{cLFC}}^{\text{max}}$ would imply a breakdown of cLFC.

Also shown in the last three columns of  TABLE~\ref{tab:brbounds} are the expected effective yields, \textit{i.e.}, yields \eqref{eq:Num_hc} divided by the reconstruction efficiency $\eta_{\text{eff}}$ for the benchmarks at Belle II and FCC-ee, the latter in parenthesis. 
With the exception of $D_s^+$-decays and $D^0 \to K^+ K^- \nu \bar \nu$, all decays give maximal expected rates $\mathcal{O}(10^6) \gtrsim N^\text{exp}/\eta_\text{eff} \gtrsim \mathcal{O}(10^4)$ at Belle II, and an order of magnitude larger for the 
FCC-ee benchmark \cite{Abada:2019lih}.
This reiterates that projected reaches at $e^+ e^-$--machines could make a $5\,\sigma$ NP discovery in different modes, and provide information
on charged lepton flavor symmetries.

\begin{table}[ht!]
 \centering
    \resizebox{0.45\textwidth}{!}{
 \begin{tabular}{lcccccc}
  \hline
  \hline
$h_c\to F $ & $\mathcal{B}_{\text{LU}}^{\text{max}}$ &  $\mathcal{B}_{\text{cLFC}}^{\text{max}}$  &  $\mathcal{B}^{\text{max}}$ & $N^\text{max}_\text{LU}/\eta_\text{eff}$ & $N^\text{max}_\text{cLFC}/\eta_\text{eff}$ & $N^\text{max}/\eta_\text{eff}$ \\
 & $[10^{-7}]$ & $[10^{-6}]$ & $[10^{-6}]$ &  &  &  \\
  \hline
  $D^0\to\pi^0$ & $6.1$ & $3.5$ & $13$ & $47\kil\,(395\kil)$ & $270\kil\,(2.3\mill)$ & $980\kil\,(8.3\mill)$ \\
  $D^+\to\pi^+$ & $25$& $14$& $52$ & $77\kil\,(650\kil)$ & $440\kil\,(3.7\mill)$ & $1.6\mill\,(14\mill)$  \\
  $D_s^+\to K^+$ & $4.6$& $2.6$& $9.6$ & $6\kil\,(50\kil)$ & $34\kil\,(290\kil)$ & $120\kil\,(1.1\mill)$  \\
  &&&& & & \\
  $D^0\to\pi^0\pi^0$ & $1.5$& $0.8$& $3.1$& $11\kil\,(95\kil)$ & $64\kil\,(540\kil)$ & $230\kil\,(2.0\mill)$\\
  $D^0\to\pi^+\pi^-$ & $2.8$& $1.6$& $5.9$& $22\kil\,(180\kil)$ & $120\kil\,(1.0\mill)$ & $450\kil\,(3.8\mill)$ \\
  $D^0\to K^+K^-$ & $0.03$& $0.02$& $0.06$ & $0.2\kil\,(1.9\kil)$ & $1.3\kil\,(11\kil)$ & $4.8\kil\,(40\kil)$ \\
  &&&&& &  \\
  $\Lambda_c^+\to p^+$ & $18$ & $11$ & $39$ & $14\kil\,(120\kil)$ & $82\kil\,(700\kil)$ & $300\kil\,(2.6\mill)$ \\
  $\Xi_c^+\to \Sigma^+$ & $36$ & $21$ & $76$ & $28\kil\,(240\kil)$ & $160\kil\,(1.4\mill)$ & $590\kil\,(5.0\mill)$ \\
  &&&&& &  \\
  {$D^0\to X$} & $15$ & $8.7$ & $32$ & $120\kil\,(980\kil)$ & $660\kil\,(5.6\mill)$ & $2.4\mill\,(21\mill)$ \\
  {$D^+\to X$} & $38$ & $22$ & $80$ & $120\kil\,(1.0\mill)$ & $680\kil\,(5.8\mill)$ & $2.5\mill\,(21\mill)$ \\
  {$D_s^+\to X$} & $18$ & $10$ & $38$ & $24\kil\,(200\kil)$ & $140\kil\,(1.1\mill)$ & $500\kil\,(4.2\mill)$ \\
  \hline
  \hline
  \end{tabular}}
  \caption{Upper limits on branching ratios $\mathcal{B}_{\text{LU}}^{\text{max}}$, $\mathcal{B}_{\text{cLFC}}^{\text{max}}$ and $\mathcal{B}^{\text{max}}$ corresponding to Eqs.~\eqref{eq:LU1}, \eqref{eq:cLFC1} and \eqref{eq:total1}, respectively, using Eq.~\eqref{eq:BR} and TABLE~\ref{tab:afactors}.
  The expected number of events \eqref{eq:Num_hc} per reconstruction efficiency $\eta_\text{eff}$ for Belle II with $50\,\text{ab}^{-1}$  \cite{Abada:2019lih} (FCC-ee yields in parentheses) corresponding to LU, cLFC, and general are displayed in the last three columns.}
    \label{tab:brbounds}
\end{table}

\subsection{Consistency checks using different modes \label{sec:corr}}

In the weak effective theory \eqref{eq:Heff_noright} only two combinations of Wilson coefficients $x_U^\pm$ describe all $h_c \to F \,\nu \bar \nu$ modes.
The system is therefore overconstrained, and allows for consistency checks. 
The sensitivity to the coefficients differs from mode to mode, as observed in Section~\ref{sec:branchingratios} from  Eq.~\eqref{eq:BR} together with items 
(a)--(d). 
In particular $D \to P \,\nu \bar \nu $ and $D \to P_1 P_2 \, \nu \bar \nu $ are essentially orthogonal to each other, 
the former depends on $x_U^+$ only, while the
latter is predominantly induced by $x_U^-$.
We can therefore predict all other branching ratios if any of these two are measured
\begin{align}\label{eq:correlations}
    \begin{split}
    \mathcal{B}(h_c\to F\,\nu\,\bar \nu)\,&=\,r_+^{h_c F}\,\mathcal{B}(D \to P \,\nu \bar{\nu}) \\
    &+\,r_-^{h_c F}\,\mathcal{B}(D^{(\prime)} \to P_1\,P_2\, \nu\, \bar{\nu})\,,
    \end{split}
\end{align}
where $r_+^{h_c F}=A^{h_c F}_+/A^{DP}_+$ and $r_-^{h_c F}=A^{h_c F}_-/A^{D P_1 P_2}_-$, up to
corrections of the order $A^{D P_1 P_2}_+/A^{D P_1 P_2}_- \lesssim 10^{-2}$.
Notice that  Eq.~\eqref{eq:correlations} holds for identical and  different $D$-mesons, $D\neq D^\prime$.
Eq.~\eqref{eq:correlations} is independent of $x_U^\pm$ and 
hence tests the assumptions that enter the effective theory framework. 
The correlation between different dineutrino modes could, for instance, be broken in the presence of additional Wilson coefficients. 
A possibility are RH light neutrinos, discussed next.

\subsection{Including light right-handed neutrinos}\label{sec:RH}

We consider going beyond the weak effective theory framework \eqref{eq:Heff_noright} by allowing for light RH neutrinos.
A concrete model with LNV is discussed in Section \ref{sec:lnv}.
With light RH neutrinos  further dimension six dineutrino operators are allowed, such as  vector and axial-vector ones,
\begin{align} \label{eq:opsplusO}
\begin{split}
Q_{LR}^{ij}&=  (\bar u_L \gamma_\mu c_L)  \,  (\bar \nu_{jR} \gamma^\mu \nu_{iR}) \,, \\
Q_{RR}^{ij} &=  (\bar u_R \gamma_\mu c_R) \,  (\bar \nu_{jR} \gamma^\mu \nu_{iR}) \,, 
\end{split}
\end{align}
and those with quark chirality mixing 
\begin{align}
\begin{split}\label{eq:opsplusOSP}
Q_{S (P)}^{i j}  &= (\bar u_{L} c_{R}) \,(\bar \nu_{j}\,(\gamma_5)\,\nu_{i}) \,, \\
Q_{T (T5)}^{i j}  &= \frac{1}{2} (\bar u \,\sigma_{\mu\nu}\, c)\, (\bar \nu_{j} \,\sigma^{\mu\nu}\,(\gamma_5)\, \nu_{i})  \,,
\end{split}
\end{align}
in addition to the chirality-flipped $Q^\prime$ operators which are obtained from the $Q$'s by interchanging left-handed ($L$) and right-handed ($R$) chiral
fields, $L\leftrightarrow R$. 
While for the SM-like neutrino case the definition of $x_U$ was useful, with light RH neutrinos it is convenient to define the following combination of Wilson coefficients, as
\begin{align}\label{eq:yU}
    y_U=\sum_{i,j}\left(\vert \mathcal{C}_S^{ij}-\mathcal{C}_{S}^{\prime ij} \vert^2+\vert \mathcal{C}_P^{ij}-\mathcal{C}_{P}^{\prime ij} \vert^2\right)\,.
\end{align} 
This particular combination enters the branching ratio of $D^0 \to \nu \bar \nu$ decays, which is constrained by 
Belle~\cite{Lai:2016uvj}
\begin{align}\label{eq:bellebound}
  \mathcal{B}(D^0 \to \text{inv.}) < 9.4 \cdot 10^{-5}\,,  
\end{align}
at 90 $\%$ CL. 
From here we obtain the constraint
\begin{align}\label{eq:boundSP}
y_U\lesssim \frac{64\,\pi^3\, m_c^2\, \mathcal{B}(D^0 \to \text{inv.})}{G_{\text{F}}^2\, \alpha_e^2\, m_D^5\, f_D^2\, \tau_D}\sim 67\,, 
\end{align}
with the decay constant $f_D=0.212\,\text{GeV}$~\cite{Aoki:2019cca}. 
Contributions from vector and axial-vector operators to $D^0\to\nu \bar \nu$
are helicity suppressed by two powers of the neutrino mass, and negligible. Tensor operators do not contribute to $D^0\to\nu \bar \nu$ decays at all. 
Only scalar and pseudoscalar operators as in $y_U$ are therefore constrained by \eqref{eq:bellebound}.

Considering either $\mathcal{C}^{ij}_{P,S}=0$ or $\mathcal{C}^{\prime ij}_{P,S}=0$, the branching ratio of $D\to P\,\nu\bar\nu$ decays,
which unlike $D^0 \to \nu\bar \nu$, depends on the sum of  $\mathcal{C}^{ij}_{P,S}$  and $\mathcal{C}^{\prime ij}_{P,S}$, can be written as
\begin{align}\label{eq:BRSP}
    \mathcal{B}(D\to P\,\nu\bar\nu)_{S,P}\,=\,A_0^{DP}\,y_U\,,
\end{align}
with
\begin{align}
    A_0^{DP}=\int_{q^2_{\text{min}}}^{q^2_{\text{max}}}\,\text{d}q^2\,a_0^{DP}(q^2)\,,
\end{align}
and
\begin{align}
    a_0^{DP}(q^2)=\frac{\tau_D\,G_{\text{F}}^2\, \alpha_e^2\, {\lambda(m_D^2,m_P^2,q^2)}^{\frac{1}{2}}}{1024\,\pi^5\, m_D^3}\nonumber\\
   \times  \frac{q^2}{m_c^2}\,(m_D^2-m_P^2)^2\, ({f_0^{DP}(q^2)})^2\,,
\end{align}
where $q^2_{\text{min}}$ and $q^2_{\text{max}}$ are the kinematic limits of $D\to P\,\nu\bar\nu$, see Section \ref{sec:DP}. 
We  provide the impact exemplarily on $D \to P\,\nu\bar\nu$ decays since there is no specific enhancement or suppression in semileptonic decays for  $S,P$-operators.
Using Eq.~\eqref{eq:BRSP} together with Eq.~\eqref{eq:boundSP}, we obtain the following limits  
\begin{align} \label{eq:SP}
    \begin{split}
        \mathcal{B}\left( D^0  \to \pi^0 \,\nu\bar{\nu}\right)_{S,P} &\lesssim \phantom{1}2.4 \cdot 10^{-6}\,,\\
        \mathcal{B}\left( D^+  \to \pi^+ \,\nu\bar{\nu}\right)_{S,P} &\lesssim 12.2  \cdot 10^{-6}\,,\\
        \mathcal{B}\left( D^+_s  \to K^+\, \nu\bar{\nu}\right)_{S,P} &\lesssim \phantom{1}2.3  \cdot 10^{-6}\,.\\
    \end{split}
\end{align}
These represent corrections of $\sim 20\,\%$ to the general flavor branching ratio limits  for $D\to P\,\nu \bar \nu$ decays  given in TABLE~\ref{tab:brbounds}.
The upper limits  based on lepton flavor conservation receive order one corrections, but the overall size of $\mathcal{B}^\text{max}_\text{cLFC}$  remains.
The upper limits based on LU are overwhelmed by \eqref{eq:SP}.

On the other hand, effects from scalar and pseudoscalar operators could become irrelevant, if an improved bound for $\mathcal{B}(D^0\to \text{inv.})$ would become available. 
Requiring the effect of $S,P$-operators on the $D\to P \,\nu\bar\nu$ branching ratios assuming  LU to be less than $\sim 10\,\%$, and thus within the uncertainties, we find $y_U\lesssim 1.7$ and
\begin{align}\label{eq:10percent2}
\mathcal{B}(D^0\to \text{inv.})&\lesssim 2\cdot10^{-6}\,.
\end{align}
An improvement of the current bound Eq.~\eqref{eq:bellebound} by two orders of magnitude as in \eqref{eq:10percent2} would exclude large scalar and pseudoscalar contributions to rare dineutrino charm decays and thus reinforce our framework and the LU limits from TABLE~\ref{tab:brbounds}.

\subsection{Bounding Lepton Number Violation \label{sec:lnv}}

Since the final states are invisibles, Eq.~\eqref{eq:bellebound} provides  opportunities to probe exotic BSM physics. 
In particular, the  final state could be two neutrinos, allowing to probe LNV in $\Delta L=2$ transitions. 
While such processes are forbidden in the SM, they occur in neutrino mass models of Majorana type.

To discuss the implication of LNV on our study we work within the standard model effective theory (SMEFT), which has already been instrumental in  Ref.~\cite{Bause:2020auq} to achieve model-independent links between  left-handed dineutrino and charged dilepton couplings, as detailed in Appendix \ref{sec:detailsu2}. 
In SMEFT higher dimensional operators consistent with Lorentz- and $SU(3)_C \times SU(2)_L \times U(1)_Y$-invariance are composed out of SM degrees of freedom. 
It is assumed that the scale of NP, here the scale of LNV, $\Lambda_\text{LNV}$, is sufficiently separated from the weak scale.

The lowest order contribution to $c \to u \,\nu \nu$ modes at tree level  is induced by a single  dimension seven operator~\cite{deGouvea:2007qla},
\begin{align}\label{eq:LNVoperator}
    \mathcal{O}^{(7)}_{4a}=L^\alpha_i\,L^\beta_j\,\bar{Q}_\alpha^b\,\bar{U}^c_a\,H^\rho\,\epsilon_{\beta\rho}\,,
\end{align}
with leptons $L=(\nu_L,\ell_L)$, quarks $Q=(u_L,d_L)$ and the Higgs $H=(H^+,H^0)$, all of which  are $SU(2)_L$-doublets, 
and the singlet up-type quarks $U$. 
Here, the superscript $c$ denotes charge conjugation and $\alpha,\beta$ are $SU(2)_L$ indices, while $i,j,a,b$ are flavor indices. 
 
Following  \cite{Deppisch:2020oyx}, we account for the different contractions between $SU(2)_L$ indices and rewrite Eq.~\eqref{eq:LNVoperator} using four-spinor notation. 
We find that $\mathcal{O}^{(7)}_{4a}$ induces contributions to the chirality flipping operators
in the weak effective hamiltonian \eqref{eq:opsplusOSP}. 
The contribution to the scalar and pseudoscalar operators reads
\begin{align}
    \mathcal{C}_{S(P)}^{ij}\,=\,\sqrt{2}\,\left(\frac{2\,\pi}{\alpha_e}\right)\,\left(\frac{v}{\Lambda_{\text{LNV}}^{ij}}\right)^3\,.
\end{align}
Here we shuffled the flavor dependence in the Wilson coefficients of $\mathcal{O}^{(7)}_{4a}$ to the one in the scale, and $v=246\,\text{GeV}$ is the SM Higgs vacuum expectation value (VEV).
There are also contributions to tensors in addition to $Q_{S,P}^\prime$ contributions but following Section \ref{sec:RH} these are not relevant to investigate the impact on the dineutrino branching ratios.
In terms of \eqref{eq:yU},
\begin{align}\label{eq:y_matching}
    y_U\,=\,4\,\left(\frac{2\,\pi}{\alpha_e}\right)^2\,  \sum_{i,j} \left(\frac{v}{\Lambda^{ij}_{\text{LNV}}}\right)^6\,.
\end{align}
Using the upper limit on $\mathcal{B}(D^0 \to \text{inv.})$ from Eq.~\eqref{eq:bellebound}, we obtain a lower limit on the LNV scale from charm, 
\begin{align} \label{eq:c-LNV}
   \Lambda^{ij}_{\text{LNV}}\gtrsim 1.5\,\text{TeV}\ .
\end{align}
This limit  is obtained assuming one term of  fixed lepton flavor  indices at a time. In the presence of more than one term the limit gets stronger.

The relation \eqref{eq:y_matching} can also be used to estimate the minimal scale $\Lambda_{\text{min}}^{ij}$  required to not spoil the results in TABLE~\ref{tab:brbounds}. 
Along the lines of the analysis in the previous Section \ref{sec:RH},
we require the branching ratio $\mathcal{B}_{\text{LU}}(D^0\to\pi^0\,\nu\bar\nu$) not to be altered by LNV contributions by more than $10\,\%$. 
We obtain
\begin{align}\label{eq:boundLNV1}
    \Lambda_{\text{min}}^{ij}&\gtrsim 2.7\,\text{TeV}\,,
\end{align}
which is much lower than the typical neutrino see-saw scale in grand unification theory (GUT) models, 
and also sufficiently below the one obtained recently from rare kaon decays, $\Lambda^{ij}_{\text{LNV}} \gtrsim 15\,\text{TeV}$~\cite{Deppisch:2020oyx}.
Additionally, one can extract information from the neutrino mass. 
Neutrino oscillations~\cite{Zyla:2020zbs}, Tritium decay~\cite{Aker:2019uuj} and cosmological data~\cite{Ade:2015xua} require the neutrinos to be lighter than $m_\nu<0.1\,\text{eV}$.  
With the quark legs closed to a loop and the Higgs fixed to its VEV the operator \eqref{eq:LNVoperator} induces 
corrections to the neutrino masses~\cite{Deppisch:2020oyx}
\begin{align} \label{eq:nu-lnv}
\delta m_\nu  \sim\frac{1}{\sqrt{2}} \frac{m_{\text{up}}}{(4 \pi)^2} \frac{v}{ \Lambda_{\text{LNV}}} \,.
\end{align}
Requiring this  to not exceed the upper limit of $m_\nu$, we obtain,  employing for $m_{\text{up}}$  the mass of the first generation up quark 
\begin{align}\label{eq::nu-lnv2}
    \Lambda_{\text{LNV}}
    \gtrsim \mathcal{O}(10^4)\,\text{TeV}\,.
\end{align}
This  would correspond to $\mathcal{B}(D^0 \to \nu \nu)_{\text{LNV}} \lesssim \mathcal{O}(10^{-29})$
and would rule out any imprint of LNV in rare charm decays.

In conclusion, observation of $D^0 \to \text{inv.}$ around \eqref{eq:bellebound} could in principle be due to LNV, with breaking scale as low as a TeV.
In addition, LNV could also affect bounds in TABLE~\ref{tab:brbounds}.
However, such effects require a high level of
flavor tuning, given other constraints, such as \eqref{eq::nu-lnv2} and the limits from $K \to \pi \,\nu \nu$.
They could be excluded altogether with an improved search for $D^0 \to \text{invisibles}$.

\section{BSM Tree-level mediators}\label{sec:tree-level}

In this section we analyze $c \to u \nu \bar \nu$ transitions in simplified BSM frameworks, based on $Z^\prime$ and LQ models, both of which are interesting as they induce charm FCNC's at tree level.
The advantage of working with specific tree-level mediators is that this circumvents the use of data on the down-sector, see  Appendix~\ref{sec:detailsu2},
allowing for a direct link between up-sector charged dilepton data and the dineutrino modes.

To be specific,
the $SU(2)_L$-links  \cite{Bause:2020auq} in SMEFT involve the leading dimension six four-fermion operators
\begin{align}\label{eq:ops} 
\begin{split}
\mathcal{L}_{\text{SMEFT}} & \supset \frac{C^{(1)}_{\ell q}}{v^2} \bar Q \gamma_\mu Q \,\bar L \gamma^\mu L +\frac{C^{(3)}_{\ell q}}{v^2} 
\bar Q \gamma_\mu  \tau^a Q \,\bar L \gamma^\mu \tau^a L   \\
&+\frac{C_{\ell u}}{v^2}  \bar U \gamma_\mu U \,\bar L \gamma^\mu L +
\frac{C_{\ell d}}{v^2} \bar D \gamma_\mu D \,\bar L \gamma^\mu L \,,
\end{split}
\end{align}
where $\tau^a$ are Pauli-matrices, while $Q$ and $L$ denote left-handed quark and lepton $SU(2)_L$--doublets, whereas $U,D$ stand for right-handed
up-singlet, down-singlet quarks,
respectively, with quark and lepton flavor indices suppressed for brevity.  
We can write the operators above in terms of its $SU(2)_L$-components and read off dineutrino Wilson coefficients $(C_A^P)$ and charged dilepton ones $(K_A^P)$ for $P=U$ ($P=D$), which refers to the up-quark sector (down-quark sector) and $A=L (R)$ denotes left- (right-) handed quark currents.
Model-independently holds
\begin{align}  \label{eq:link}
\begin{split}
    C_L^U&=K_L^D=C^{(1)}_{\ell q} + C^{(3)}_{\ell q}  \,, \quad C_R^U=K_R^U=C_{\ell u}  \, , \\
    C_L^D&=K_L^U=C^{(1)}_{\ell q}  - C^{(3)}_{\ell q}  \, , \quad
    C_R^D=K_R^D=C_{\ell d}  \, .   
\end{split}
\end{align}
While $C_R^P=K_R^P$, due to the different relative signs between $C^{(1)}_{\ell q}$ and $C^{(3)}_{\ell q}$, the left-handed dineutrino  couplings relevant for charm, $C_L^U$,  are linked to the down-sector dilepton ones, $K_L^D$, and require hence input from strange quarks.
BSM models with tree level mediators, in which the relation between $C^{(1)}_{\ell q}$ and $C^{(3)}_{\ell q}$ is known, are simpler.
Specifically, we study models with 
\begin{align} \label{eq:frame}
 C_{\ell q}^{(3)}=
  \begin{cases}
  ~0~, & \text{$Z^\prime$ models~,} \\[0.25cm]
    ~\alpha\,C_{\ell q}^{(1)}~, & \text{LQ models~.} 
  \end{cases}
\end{align}
Values of $\alpha$ for different LQ representations are given in TABLE~\ref{tab:LQ}. 
In the following, we work out the upper limits on the dineutrino  branching ratios assuming \eqref{eq:frame}.
We also consider LQs induced by right-handed operators with $C_{\ell q}^{(1,3)}=0$.
The  results are displayed in TABLES~\ref{tab:brboundsNP_Zprime} and \ref{tab:brboundsNP_LQ}.
We stress that our results correspond to quite generic BSM frameworks:
the sole "model-dependent" input we use is the matching condition \eqref{eq:frame}.

\subsection{$\boldsymbol{Z^\prime}$ models}\label{sec:zprime}

In $Z^\prime$ models, the following link between dineutrino ($C$) and charged lepton ($K$) Wilson coefficients in the gauge basis holds
\begin{align}\label{eq:restrictionZp}
    C_L^U=K_L^U=C_{\ell q}^{(1)}\,.
\end{align}
From Eq.~\eqref{eq:super} follows
\begin{align}
    x_U^{Z^\prime}<\sum_{i,j} \left( \big\vert\mathcal{K}_{R}^{U_{12}\, ij}\big\vert^2  +    \big\vert\mathcal{K}_{L}^{U_{12} ij}\big\vert^2  \right)\,,
\end{align}
where bounds  on the couplings on the right hand side can be seen in TABLE~\ref{tab:limitsonK}.
We obtain
\begin{align}  \label{eq:ZpLU}
x_U^{Z^\prime} &\lesssim 15 \,, \quad  (\text{LU}) \\ \label{eq:ZpcLFC}
x_U^{Z^\prime} & \lesssim 85 \,, \quad (\text{cLFC}) \\ \label{eq:Zptotal}
x_U^{Z^\prime} &  \lesssim 288 \, ,  \quad (\text{general}) \,, 
\end{align}
which are  stronger than the model-independent ones \eqref{eq:LU1}-\eqref{eq:total1}, however, within  the same order of magnitude.
Upper limits on dineutrino branching ratios from \eqref{eq:restrictionZp} are given in TABLE~\ref{tab:brboundsNP_Zprime}.
\begin{table}[ht!]
 \centering
 \begin{tabular}{l|ccc}
  \hline
  \hline
  & \multicolumn{3}{c}{$Z^\prime$} \\ 
  \hline
  $h_c\to F$ & $\mathcal{B}_{\text{LU}}^{\text{max}}$ &  $\mathcal{B}_{\text{cLFC}}^{\text{max}}$  &   $\mathcal{B}^{\text{max}}$  \\
  & $[10^{-7}]$ & $[10^{-6}]$ & $[10^{-6}]$ \\
  \hline
  $D^0\to\pi^0$ & $2.7$ & $1.5$ & $5.1$ \\ 
  $D^+\to\pi^+$ & $11$ & $6.1$ & $21$ \\ 
  $D_s^+\to K^+$& $2.0$ & $1.1$ & $3.9$ \\ 
  & & &  \\ 
  $D^0\to\pi^0\pi^0$ & $0.6$ & $0.4$ & $1.2$ \\ 
  $D^0\to\pi^+\pi^-$ & $1.2$ & $0.7$ & $2.4$ \\ 
  $D^0\to K^+K^-$ & $0.01$ & $0.007$ & $0.03$ \\ 
  & & &  \\ 
  $\Lambda_c^+\to p^+$ & $8.0$ & $4.6$ & $16$ \\ 
  $\Xi_c^+\to \Sigma^+$ & $16$ & $9.0$ & $31$ \\ 
  & & &  \\ 
  $D^0\to X$ & $6.6$ & $3.8$ & $13$ \\ 
  $D^+\to X$ & $17$ & $9.5$ & $32$ \\ 
  $D_s^+\to X$ & $7.9$ & $4.5$ & $15$ \\ 
  \hline
  \hline
  \end{tabular}
  \caption{$\mathcal{B}_{\text{LU}}^{\text{max}}$, $\mathcal{B}_{\text{cLFC}}^{\text{max}}$ and $\mathcal{B}^{\text{max}}$  corresponding to the LU, cLFC and general bounds, respectively, for $Z^\prime$ models \eqref{eq:restrictionZp}, using Eq.~\eqref{eq:BR} together with the values of $A_\pm^{h_c\,F}$ displayed in TABLE~\ref{tab:afactors} and the bounds on $x_U$ from Eqs.~\eqref{eq:ZpLU}-\eqref{eq:Zptotal}.
  }
  \label{tab:brboundsNP_Zprime}
\end{table}

\subsection{Leptoquark models}\label{sec:LQbounds}

In contrast to $Z^\prime$ models, in LQ models  both Wilson coefficients $C^{(1)}_{\ell q}$ and $C^{(3)}_{\ell q}$ can contribute.
The general charged lepton-neutrino link reads
\begin{align}\label{eq:LQrelations}
\begin{split}
    K_L^U=\gamma\,C_L^U=(1-\alpha)\,C_{\ell q}^{(1)}\,, ~~\gamma=\frac{1-\alpha}{1+\alpha} \, . 
\end{split}
\end{align}
The values of $\alpha$ for different LQ models are given in TABLE~\ref{tab:LQ}. 
Here, $V, S$ denote vector, scalar LQs and the subscript indicates the dimension of the representation under $SU(2)_L$.
Representations with tree level contribution to $c \to u \nu \bar \nu$ are the triplets $S_3,V_3$, the doublets
$S_2, \tilde V_2$, and the singlet $V_1$. 
We discuss them separately in the following.

\addtolength{\tabcolsep}{2pt} 
\begin{table}[h!]
 \centering
 \begin{tabular}{ccccc}
  \hline
  \hline
  LQ-rep & $\alpha$ & $\gamma$ & $C_{L,R}^{U}$  \\
  \hline
  $V_1$ &$\phantom{-}1$ & $0$ & $C_R^U=0$  \\
  $S_3$ &$\phantom{-}\frac{1}{3}$ & $\frac{1}{2}$ &  $C_R^U=0$  \\
  $V_3$ &$-\frac{1}{3}$& $2$ & $C_R^U=0$  \\ 
  \hline
  $S_2, \tilde V_2$ & -- & -- &  $C_L^U=0$ \\
  \hline
  \hline
  \end{tabular}
  \caption{Values for $\alpha$ and $\gamma$ \eqref{eq:frame}, \eqref{eq:LQrelations} for different LQ representations~\cite{Hiller:2016kry}. The last column displays which dineutrino Wilson coefficient is not generated by the LQ representation.}
\label{tab:LQ}
\end{table}

\begin{table*}[ht!]
 \centering
     \resizebox{0.9\textwidth}{!}{
 \begin{tabular}{l|ccc|ccc|ccc|ccc}
  \hline
  \hline
 \mbox{} & \multicolumn{3}{c}{$S_3$}  & \multicolumn{3}{c}{$V_3$} & \multicolumn{3}{c}{$S_2$,\,$\tilde{V}_2$} & \multicolumn{3}{c}{$V_1$} \\ 
  \hline
  $h_c\to F$ & $\mathcal{B}_{\text{LU}}^{\text{max}}$ &  $\mathcal{B}_{\text{cLFC}}^{\text{max}}$ & $\mathcal{B}^{\text{max}}$ & $\mathcal{B}_{\text{LU}}^{\text{max}}$ &  $\mathcal{B}_{\text{cLFC}}^{\text{max}}$ & $\mathcal{B}^{\text{max}}$ & $\mathcal{B}_{\text{LU}}^{\text{max}}$ &  $\mathcal{B}_{\text{cLFC}}^{\text{max}}$ & $\mathcal{B}^{\text{max}}$ & $\mathcal{B}_{\text{LU}}^{\text{max}}$ &  $\mathcal{B}_{\text{cLFC}}^{\text{max}}$ & $\mathcal{B}^{\text{max}}$\\
  & $[10^{-7}]$ & $[10^{-6}]$ & $[10^{-6}]$ & $[10^{-7}]$ & $[10^{-6}]$ & $[10^{-6}]$ & $[10^{-7}]$ & $[10^{-6}]$ & $[10^{-6}]$ & $[10^{-7}]$ & $[10^{-6}]$ & $[10^{-6}]$\\
  \hline
  $D^0\to \pi ^0$ & 5.3 & 3.0 & 10 & 0.3 & 0.2 & 0.6 & 1.3 & 0.8 & 2.6 & 4.8 & 2.7 & 10 \\
 $D^+\to \pi ^+$ & 21 & 12 & 42 & 1.3 & 0.8 & 2.6 & 5.3 & 3.1 & 10 & 19 & 11 & 41 \\
 $D_s^+\to K^+$ & 4.0 & 2.3 & 7.7 & 0.2 & 0.1 & 0.5 & 1.0 & 0.6 & 1.9 & 3.6 & 2.1 & 7.7 \\
 &&& &&& &&& &&& \\
 $D^0\to \pi ^0 \pi ^0$ & 1.3 & 0.7 & 2.5 & 0.08 & 0.05 & 0.2 & 0.3 & 0.2 & 0.6 & 1.1 & 0.7 & 2.4 \\
 $D^0\to \pi ^+ \pi ^-$ & 2.4 & 1.4 & 4.8 & 0.2 & 0.09 & 0.3 & 0.6 & 0.4 & 1.2 & 2.2 & 1.3 & 4.7 \\
 $D^0\to K^+ K^-$ & 0.03 & 0.01 & 0.05 & 0.002 & 0.001 & 0.003 & 0.006 & 0.004 & 0.01 & 0.02 & 0.01 &
   0.05 \\
  &&& &&& &&& &&& \\  
 $\Lambda _c^+\to p^+ $& 16 & 9.2 & 31 & 1.0 & 0.6 & 1.9 & 4.0 & 2.3 & 7.8 & 14 & 8.3 &
   31 \\
$\Xi _c^+\to \Sigma^+$ & 31 & 18 & 61 & 2.0 & 1.1 & 3.8 & 7.9 & 4.5 & 15 & 28 & 16
   & 61\\
 &&& &&& &&& &&& \\ 
$D^0\to X$ & 13 & 7.5 & 25 & 0.8 & 0.5 & 1.6 & 3.3 & 1.9 & 6.4 & 12 & 6.8 & 25 \\
$D^+\to X$ & 33 & 19 & 65 & 2.1 & 1.2 & 4.0 & 8.3 & 4.8 & 16 & 30 & 17 & 64 \\
$D_s^+\to X$ & 16 & 9.0 & 31 & 1.0 & 0.6 & 1.9 & 3.9 & 2.3 & 7.7 & 14 & 8.2 & 30 \\
  \hline
  \hline
  \end{tabular}}
\caption{$\mathcal{B}_{\text{LU}}^{\text{max}}$, $\mathcal{B}_{\text{cLFC}}^{\text{max}}$ and $\mathcal{B}^{\text{max}}$  corresponding to the LU, cLFC and general bounds, respectively, for LQ models using Eq.~\eqref{eq:BR} together with the values of $A_\pm^{h_c\,F}$ displayed in TABLE~\ref{tab:afactors} and the bounds on $x_U$ determined in Section~\ref{sec:LQbounds}.
}
\label{tab:brboundsNP_LQ}
\end{table*}

\subsubsection{Triplets $S_3$ and $V_3$ }

The dineutrino contributions from the LQ representations $S_3$ and $V_3$ are related to the charged dilepton bounds as
\begin{align}
    x_U^{S_3,\,V_3}<\frac{1}{|\gamma|^2}\,\sum_{i,j} \big\vert\mathcal{K}_{L}^{U_{12} ij}\big\vert^2\,,
\end{align} 
and right-handed contributions are absent.  
For the scalar triplet  we obtain using  TABLE~\ref{tab:limitsonK}
\begin{align}\label{eq:S3LU}
x_U^{S_3} &\lesssim 30 \,, \quad  (\text{LU}) \\ \label{eq:S3LFC}
x_U^{S_3} & \lesssim 170 \,, \quad (\text{cLFC}) \\ \label{eq:S3total}
x_U^{S_3} &  \lesssim  577 \,\quad (\text{general}) \,, 
\end{align}
whereas the ones for the $V_3$ are a factor $1/16$ smaller (modulo rounding effects) and read  
\begin{align} \label{eq:LQV3LU}
x_U^{V_3} &\lesssim  2\,, \quad  (\text{LU}) \\ \label{eq:LQV3cLFC}
x_U^{V_3} & \lesssim 11\,, \quad (\text{cLFC}) \\ \label{eq:LQV3total}
x_U^{V_3} &  \lesssim 36 \, ,  \quad (\text{general}) \,.
\end{align}

\subsubsection{Doublets  $S_2$ and $\tilde{V}_2$}

The doublet LQs induce right-handed contributions only, as
\begin{align}
    x_U^{S_2,\tilde{V}_2}<\sum_{i,j}   \big\vert\mathcal{K}_{R}^{U_{12} ij}\big\vert^2\,.
\end{align}
Using TABLE~\ref{tab:limitsonK}, we obtain for both scalar and vector representations
\begin{align}  \label{eq:LQ4LU}
x_U^{S_2,\tilde{V}_2} &\lesssim 7\,, \quad  (\text{LU}) \\ \label{eq:LQ4cLFC}
x_U^{S_2,\tilde{V}_2} & \lesssim 42\,, \quad (\text{cLFC}) \\ \label{eq:LQ4total}
x_U^{S_2,\tilde{V}_2} &  \lesssim  144 \, ,  \quad (\text{general}) \,,
\end{align}
which is, modulo rounding errors, half of the ones in the $Z^\prime$ model given in \eqref{eq:ZpLU}-\eqref{eq:Zptotal}.
The reason for this is that the constraints obtained from Drell-Yan processes do not depend on the quark current chirality.

\subsubsection{Singlet $V_1$}

In the LQ representation $V_1$ right-handed currents are absent, and $\mathcal{K}_L^{U}=0$.
Hence, no connection between $\mathcal{K}_L^{U}$ and $\mathcal{C}_L^{U}$ exists. 
However, model-independently $\mathcal{C}_L^{U}=\mathcal{K}_L^{D}$ \eqref{eq:link}, and one can employ data from the down-sector.
Using Eq.~\eqref{eq:super}, we obtain the following bound
\begin{align}
    x_U^{V_1}<\sum_{i,j}   \big\vert\mathcal{K}_{L}^{D_{12} ij}\big\vert^2\,+\,\delta x_{U}^{V_1}\,,
\end{align}
with the linear correction from CKM-rotation at the order $\lambda$,
\begin{align}
    \delta x_{U}^{V_1}=2\,\lambda\,\sum_{i,j} \big\vert \mathcal{K}_{L}^{D_{12} ij}\big\vert \left(\big\vert \mathcal{K}_{L}^{D_{22} ij}\big\vert+\big\vert \mathcal{K}_{L}^{D_{11} ij}\big\vert\right)\,.
\end{align}
Using TABLE~\ref{tab:limitsonK}, we obtain
\begin{align} \label{eq:LQ1LU}
x_U^{V_1} &\lesssim 27\,, \quad  (\text{LU}) \\ \label{eq:LQ1cLFC}
x_U^{V_1} & \lesssim 153\,, \quad (\text{cLFC}) \\ \label{eq:LQ1total}
x_U^{V_1} &  \lesssim  572 \, ,  \quad (\text{general}) \,.
\end{align}
The LU bound in Eq.~\eqref{eq:LQ1LU} can be significantly improved to $x_U^{V_1} \lesssim \mathcal{O}(10^{-3})$ if low energy kaon data is applied. See also the discussion in Appendix~\ref{sec:detailsu2}.

\subsection{Synopsis  tree-level mediators}

In TABLES~\ref{tab:brboundsNP_Zprime} and \ref{tab:brboundsNP_LQ} 
we show the limits on the branching ratios $\mathcal{B}_{\text{LU}}^{\text{max}}$,\,$\mathcal{B}_{\text{cLFC}}^{\text{max}}$ and $\mathcal{B}^{\text{max}}$ imposing LU, cLFC and general flavor structure, respectively, 
for the $Z^\prime$ and the LQ models. We stress that our results hold for any BSM model with the same relations between
$C^{(1)}_{\ell q}$ and $C^{(3)}_{\ell q}$  and corresponding right-handed contributions, and are therefore still quite generic.
As expected,  bounds for all simplified  models are stronger than the model-independent ones shown in TABLE~\ref{tab:brbounds}.
The $Z^\prime$, $S_2,\,\tilde{V}_2$ and especially the $V_3$ are significantly better constrained, 
whereas $S_3$ and $V_1$ almost saturate the model-independent bounds.

\section{Conclusions}\label{sec:con}

We performed a comprehensive analysis of $c \to u \,\nu \bar \nu$ induced decays.
We systematically analyzed exclusive decays of $D^0,D^+,D_s^+$-mesons and
$\Lambda_c^+, \Xi_c^+$-baryons using most recent determinations of form factors, in addition to inclusive modes.
The dineutrino decays are important as they complement searches for NP with radiative and dileptonic modes, 
while being significantly cleaner than the latter from the theory point of view due to the absence of irreducible resonance backgrounds.

There is presently no experimental limit on any of the  $h_c \to F\, \nu \bar \nu$ branching ratios available,
despite the fact that all of them are clean null tests of the SM. 
Hence, any observation within foreseeable sensitivity means NP, and NP can be just around the corner.
Specifically, model-independent upper limits on branching ratios, obtained  using $SU(2)_L$ and existing bounds on charged lepton modes,
allow for upper limits as large as few $\times 10^{-5}$, see TABLE~\ref{tab:brbounds}.

Moreover, the measurements of dineutrino branching ratios constitute tests of charged lepton flavor, 
specifically, lepton-universality and charged lepton flavor conservation -- a stunning opportunity given the fact that the neutrino flavors are not reconstructed.
Branching ratios assuming charged lepton flavor conservation can be as large as $10^{-5}$, 
those in the limit of lepton universality reach few $\times 10^{-6}$.  
These limits are data-driven and will go down if improved bounds from charged leptons become available~\cite{Bause:2020auq}.
 
Furthermore, we analyzed the branching ratios in BSM models that induce rare charm dineutrino decays at tree level, leptoquarks and flavorful $Z^\prime$ models.
We find that upper limits on the branching ratios are smaller than the model-independent ones, see TABLES~\ref{tab:brboundsNP_Zprime} and \ref{tab:brboundsNP_LQ}, 
yet in the same ballpark as the model-independent ones except for the vector triplet representation $V_3$, which gives an order of magnitude lower limits.

We add that there is the possibility that the modes are observed above their upper limits given in TABLE~\ref{tab:brbounds}: 
this would signal not only NP, but NP with additional light degrees of freedom. 
An example are light right-handed neutrinos from a TeV-ish scale of LNV. 
While studies in other sectors give higher scales, it is conceivable that LNV breaking is strongly flavor-dependent.
This could be further investigated with an improved bound on $D^0 \to $ invisibles.
Further study is beyond the scope of this work.

FIG.~\ref{fig:relativebranchingDdecays} summarizes the  sensitivity to rare charm decays at a clean, high luminosity $e^+ e^-$ flavor facility such as Belle II  and FCC-ee running at the $Z$.
In view of the large  charm luminosities, and the complementarity to charged lepton probes of lepton flavor universality and conservation, we strongly encourage experimental searches for dineutrino modes.

\acknowledgments
 
We are grateful to Emi Kou and Stephane Monteil for useful communication. 
This work is supported by the \textit{Studienstiftung des Deutschen Volkes} (MG) and the \textit{Bundesministerium f\"ur Bildung und Forschung} -- BMBF (HG).

\appendix
\section{Charged lepton and neutrino links via $\boldsymbol{SU(2)_L}$--symmetry}\label{sec:detailsu2}

It was recently shown~\cite{Bause:2020auq} that $SU(2)_L$--symmetry links processes into charged dileptons with those into dineutrinos. 
Here we provide details to make this paper self-contained.

To connect the low-energy effective Hamiltonian for dineutrino transitions \eqref{eq:Heff_noright} with a charged lepton one
\eqref{eq:Heff_opsCH} in a model-independent way, it is necessary to introduce the $SU(3)_C\times SU(2)_L \times U(1)_Y$-invariant effective theory with semileptonic (axial-) vector four-fermion operators. 
At leading order~\cite{Grzadkowski:2010es}, only four operators contribute, which are given by \eqref{eq:ops}.
Tree level contributions to dineutrino modes are also induced by $Z$-penguins from dimension six 
operators with two Higgs fields and a covariant derivative.
These are constrained by electroweak and top observables, or mixing~\cite{Efrati:2015eaa,Brivio:2019ius}, and subleading.

Writing \eqref{eq:ops} in terms of mass eigenstates, that is, $Q_\alpha=(u_{L \alpha},V_{\alpha \beta} \,d_{L \beta})$ and $L_i=(\nu_{Li} , W_{ki}^* \,\ell_{L k})$ with the Cabibbo-Kobayashi-Maskawa (CKM) and Pontecorvo-Maki-Nakagawa-Sakata (PMNS) matrices, $V$ and $W$, respectively, and matching onto  Eqs.~\eqref{eq:Heff_noright} and \eqref{eq:Heff_opsCH}, the dineutrino Wilson coefficients in the up-sector, $\mathcal{C}_{L,R}^U$, read
\begin{align}
\begin{split}
        \mathcal{C}_{L}^U &=W^\dagger [V\,\mathcal{K}_{L}^D\,V^\dagger]\, W\,,\\ 
        \mathcal{C}_{R}^U&=W^\dagger [\mathcal{K}_{R}^U]\, W\,.
\end{split}
\end{align}
The $\mathcal{C}_{L,R}^U$ depend on the PMNS matrix, which drops out in the flavor-summed branching ratios \eqref{eq:sum} due to unitarity.
$\mathcal{C}_{L}^U$ depends on the CKM-matrix that allows for an expansion in the Wolfenstein parameter $\lambda$, relevant for $c \to u$ transitions as 
\begin{align*}
        \mathcal{C}_{L}^{U_{12}}=W^\dagger \mathcal{K}_{L}^{D_{12}}W\hspace{-0.1cm}+\lambda \,W^\dagger(\mathcal{K}_{L}^{D_{22}}-\mathcal{K}_{L}^{D_{11}}) W+\mathcal{O}(\lambda^2)\,.
\end{align*}
The superscripts $12,\, 11$ and $22$ given explicitly indicate the  generations in the quark currents of the  operators, \textit{i.e.},~$\bar u c,\, \bar{d} s,\, \bar{d} d$ and $\bar ss$.
In the remainder of this work, which focuses on $c \to u$ transitions, we use $\mathcal{C}_{L,R}^{U_{12}}= \mathcal{C}_{L,R}^U$ to avoid clutter.
For $x_U$, one obtains
\begin{align}
    &x_U=\sum_{\nu=i,j} \left( \big\vert\mathcal{C}_L^{U ij}\big\vert^2+\big\vert\mathcal{C}_R^{U ij}\big\vert^2 \right)=\trace \! \left[\mathcal{C}_L^U\,\mathcal{C}_L^{U \dagger}\,+\,\mathcal{C}_R^U\,\mathcal{C}_R^{U \dagger}\right]\nonumber\\
    &=  \trace \! \left[ \mathcal{K}_L^{D_{12}}{\mathcal{K}_L^{D_{12}}}^\dagger+\mathcal{K}^{U_{12}}_R {\mathcal{K}_R^{U_{12}}}^\dagger\right]
    +\, \delta x_U  +{\mathcal{O}}(\lambda^2) \nonumber\\ 
    &=\hspace{-0.2cm}\sum_{\ell=i,j} \left(   \big\vert\mathcal{K}_L^{D_{12} ij}\big\vert^2+\big\vert\mathcal{K}_R^{U_{12} ij}\big\vert^2 \right)+\, \delta x_U +{\mathcal{O}}(\lambda^2)\,,\label{eq:super} 
\end{align}
with the $\mathcal{O}(\lambda)$-correction
\begin{align}
   \delta  x_U \,&=\,2\,\lambda\,\trace\! \left[\Re{ \mathcal{K}_L^{D_{12}}\,\bigg({\mathcal{K}_{L}^{D_{22}}}^\dagger-{\mathcal{K}_{L}^{D_{11}}}^\dagger\bigg)} \right]\\
    &=2\,\lambda\,\sum_{\ell=i,j}\,\Re{ \bigg(\mathcal{K}_{L}^{D_{12}ij}\,{\mathcal{K}_{L}^{D_{22}ij}}^\ast-{\mathcal{K}_{L}^{D_{12}ij}}{\mathcal{K}_{L}^{D_{11}ij}}^\ast\bigg)}\,.\nonumber
\end{align}
The traces are over the lepton flavor indices of the Wilson coefficients, and therefore depend on the
flavor structure of the couplings $\mathcal{K}_{L,R}^{ij}$~\cite{Bause:2020auq}, see also \eqref{eq:patterns}:
\begin{enumerate}[label=(\roman*)]
\item $\mathcal{K}_{L,R}^{ij} \propto \delta_{ij}$ that is, \emph{lepton-universality} (LU).
\item $\mathcal{K}_{L,R}^{ij}$ are diagonal, that is, \emph{charged lepton flavor conservation} (cLFC)
\item $\mathcal{K}_{L,R}^{ij}$ is arbitrary.
\end{enumerate}

Interestingly, \eqref{eq:super} allows both the study of the lepton flavor nature and to put constraints on rare charm dineutrino branching ratios model-independently.
We define
\begin{align}
R^{\ell \ell^\prime}&=  \big\vert\mathcal{K}_L^{ D_{12} \ell\ell^\prime}\big\vert^2+ \big\vert\mathcal{K}_R^{U_{12} \ell\ell^\prime}\big\vert^2\,,\nonumber\\
R_\pm^{\ell \ell^\prime}&= \big\vert\mathcal{K}_L^{ D_{12} \ell\ell^\prime} \pm \mathcal{K}_R^{U_{12} \ell\ell^\prime}\big\vert^2 \, ,\label{eq:Rell}\\
\delta R^{\ell\ell^\prime} &=2\,\lambda\, \Re{\mathcal{K}_{L}^{D_{12}\ell\ell^\prime}\,{\mathcal{K}_{L}^{D_{22}\ell\ell^\prime}}^\ast-{\mathcal{K}_{L}^{D_{12}\ell\ell^\prime}}{\mathcal{K}_{L}^{D_{11}\ell\ell^\prime}}^\ast}\,.\nonumber
\end{align}
where
$R_+^{\ell\ell^\prime}+R_-^{\ell\ell^\prime}=2\, R^{\ell\ell^\prime}$,  $R_\pm^{\ell\ell^\prime} \leq 2 \,R^{\ell\ell^\prime}$. 
Furthermore, $\delta R^{\ell\ell^\prime}<2\,\lambda\,\big\vert \mathcal{K}_{L}^{D_{12} \ell \ell^\prime}\big\vert \left(\big\vert \mathcal{K}_{L}^{D_{22} \ell\ell^\prime}\big\vert+\big\vert \mathcal{K}_{L}^{D_{11} \ell\ell^\prime}\big\vert\right)$. 
We employ  high-$p_T$ data~\cite{Fuentes-Martin:2020lea,Angelescu:2020uug} for up- and down-type charged lepton FCNC's  
and give bounds on lepton specific Wilson coefficients for $\ell, \ell^\prime =e, \mu, \tau$
in TABLE~\ref{tab:limitsonK}~\footnote{The $d\to d,\,s \to s,\, s\to d$ entries can be obtained from the $c\to u$ bounds via luminosity ratios, see Eqs.~(6.9) and (6.10) in~\cite{Fuentes-Martin:2020lea} and Fig.~1 in~\cite{Angelescu:2020uug}.}. 
Corresponding  bounds on ${R^{\ell \ell^\prime}}$ and ${\delta R^{\ell\ell^\prime}}$ are 
summarized in TABLE~\ref{tab:limitsonR}.

\begin{table}[h!]
    \centering
    \begin{tabular}{c|c|cccccc}
    \hline
    \hline
     $q_i\to q_j$& $\big\vert\mathcal{K}_{A}^{P_{ji}\ell\ell^\prime}\big\vert$ & $ee$ & $\mu \mu $ & $\tau \tau$  & $e \mu$ & $e \tau$ & $\mu \tau$\\
   \hline
   $d \to d$   &$\big\vert  \mathcal{ K}_{L,R}^{D_{11} \ell \ell'} \big\vert$& $2.8$ & $1.5$ & $5.5$ & $1.1$ & $3.3$  & $3.6$ \\
   \hline
   $s \to s$ &$\big\vert  \mathcal{ K}_{L,R}^{D_{22} \ell \ell'} \big\vert$& $9.0$ & $4.9$   & $17$  & $5.2$ & $17$  & $18$ \\
   \hline
   $s \to d$ &$\big\vert  \mathcal{ K}_{L,R}^{D_{12} \ell \ell'} \big\vert$& $3.5$ & $1.9$  & $6.7$ & $2.0$ & $6.1$ & $6.6$ \\
      \hline
  $c \to u$  & $\big\vert \mathcal{ K}_{L,R}^{U_{12} \ell \ell'} \big\vert$ &   $2.9$ & $1.6$ & $5.6$ & $1.6$ & $4.7$ & $5.1$ \\
    \hline
     \hline
    \end{tabular}
    \caption{Upper limits on $|\Delta d|=0,1$ and $|\Delta c|=1$ leptonic couplings $\mathcal{ K}_{L,R}$ from  high--$p_T$~\cite{Fuentes-Martin:2020lea, Angelescu:2020uug}. 
    LFV-bounds are quoted as charge-averaged, $\sqrt{ |\mathcal{ K}^{\ell^+ \ell^{\prime-}}|^2 + |\mathcal{ K}^{\ell^- \ell^{\prime +}}|^2}$.}
    \label{tab:limitsonK}
\end{table}

\addtolength{\tabcolsep}{2pt} 
\begin{table}[h!]
    \centering
    \begin{tabular}{c|cccccc}
    \hline
    \hline
     & $ee$ & $\mu \mu $ & $\tau \tau$  & $e \mu$ & $e \tau$ & $\mu \tau$\\
    \hline
     $R^{\ell \ell'}$& $21$ & $6.0$ & $77$ & $6.6$ & $59$ & $70$ \\
     $\delta R^{\ell \ell'}$& $19$ & $5.4$ & $69$ & $5.7$& $55$ & $63$\\
     \hline
     $r^{\ell \ell'}$& $39$ & $11$ & $145$ & $12$ & $115$ & $133$ \\  
     \hline
     \hline
    \end{tabular}
    \caption{Bounds on $R^{\ell \ell'}$ and $\delta R^{\ell \ell'}$ from Eqs.~\eqref{eq:Rell},  as well as their sum,  $r^{\ell\ell^\prime}=R^{\ell \ell'}+\delta R^{\ell \ell'}$.}
    \label{tab:limitsonR}
\end{table}
\addtolength{\tabcolsep}{-2pt} 

We obtain the upper limits for the flavor patterns \eqref{eq:patterns} and general flavor structure as
\begin{align}\label{eq:LU}
x_U &= 3\,r^{\mu \mu} \lesssim 34\,, \quad  (\text{LU}) \\ \label{eq:cLFC}
x_U &= r^{e e}+ r^{\mu \mu }+r^{\tau \tau}  \lesssim 196 \,, \quad (\text{cLFC}) \\ \label{eq:total}
x_U &= r^{ee} + r^{\mu \mu} \! +r^{\tau \tau }  \!+2\, ( r^{e \mu}+ r^{e \tau}+r^{ \mu \tau} )  \lesssim  716 \,,
\end{align}
identical to \eqref{eq:LU1}-\eqref{eq:total1} and with flavor budget displayed.
Since the dimuon bounds are the most stringent ones, see TABLE~\ref{tab:limitsonR}, they provide the LU-limit.

Bounds on $\mathcal{ K}_{L,R}^{D_{ji}\ell{\ell^\prime}}$ from rare kaon decays can be stronger by about two orders magnitude than the high-$p_T$ limits for  $\ell \ell^{(\prime)}=ee$, $\mu\mu$ and $e\mu$. 
Corresponding limits on $x_U$ would be reduced to $22\,\%$  (LU), and only to $80\,\%$ (cLFC) and $92\,\%$ (general) of the ones presented in Eqs.~\eqref{eq:LU}-\eqref{eq:total}.
The latter two are dominated by contributions including $\tau$'s. Additional constraints from $\tau$-decays~\cite{Angelescu:2020uug} could be taken into account  but require further study of correlations which is beyond the scope of this work. 
Since the right-handed bounds from $c\to u$ in TABLE~\ref{tab:limitsonK} remain model-independently, the $x_U$ bounds can at most be reduced to the ones provided in Eqs.~\eqref{eq:LQ4LU}-\eqref{eq:LQ4total}. To also avoid the possibility of cancellations altogether and to use a unified framework, we therefore
present results using high--$p_T$ bounds.

\section{Parametrization of form factors}\label{app:FF}

In this appendix we provide detailed information on the form factors used in this work.
\subsection{Form factors $\boldsymbol{D \to P}$}\label{app:ff_DP}

The form factors $f_{+,0}$  for $D\to P$ are available from lattice QCD~\cite{Lubicz:2017syv}, given in the  $z$--expansion as, $i=+,0$,
\begin{align}
\label{FF_para}
\begin{split}
f_i(q^2) &= \frac{1}{1-P_i\,q^2} \biggl[ f_i(0) \\
&\quad + c_i\, (z(q^2)-z(0)) \biggl( 1 + \frac{z(q^2) + z(0)}{2} \biggr) \biggr]\,,
\end{split}
\end{align}
where
\begin{align}\label{eq:zexpansion}
\begin{split}
z(q^2) &= {\sqrt{t_+-q^2} - \sqrt{t_+-t_0} \over \sqrt{t_+-q^2} + \sqrt{t_+-t_0}}\,, 
\end{split}
\end{align}
with $t_0 = (m_D+m_P)(\sqrt{m_D}-\sqrt{m_P})^2$ and $t_+ =(m_D + m_P)^2$. 
The numerical values of $f_i(0)$, $c_i$ and $P_i$ parameters together with their uncertainties and covariance matrices are given in~\cite{Lubicz:2017syv}. 
We use the same numerical inputs for $D\to \pi$ and $D_s^+\to K^+$ transitions besides obvious kinematic replacements, supported by Ref.~\cite{Koponen:2012di}.
There is an additional factor of $1/\sqrt{2}$ for the $D^0 \to \pi^0$ form factors $f_i(q^2)$ due to isospin.

\subsection{Form factors $\boldsymbol{D \to P_1 P_2}$}\label{app:ff_DP1P2}

The transversity form factors $\mathcal{F}_i$ with $i=0,\,\perp,\,\parallel$ can be expressed in terms of three heavy hadron chiral perturbation theory (HH$\chi$PT) form factors $\omega_\pm$ and $h$ as~\cite{deBoer:2018buv,Lee:1992ih}
\begin{align}
\begin{split}\label{eq:formfactorsDPP}
\mathcal{F}_0&=\frac{\mathcal{N}_{\text{nr}}}{2}\,\Bigg[\sqrt{\lambda}\,\omega_{+} + \frac{\omega_{-}}{p^2}\left[ (m_{P_1}^2-m_{P_2}^2)\sqrt{\lambda} \right.   \\
&\hspace{1.5cm}\left.-(m_D^2-q^2-p^2)\sqrt{\lambda_{p}}\,\cos\theta_{P_1} \right] \Bigg]\,,\\
\mathcal{F}_{\parallel}&=\mathcal{N}_{\text{nr}}\,\sqrt{\lambda_p\,\frac{q^2}{p^2}}\,\omega_{-}\,,\hspace{0.3cm}
\mathcal{F}_{\perp}= \frac{\mathcal{N}_{\text{nr}}}{2}\sqrt{\lambda\lambda_p\,\frac{q^2}{p^2}}\,h  \,,\\
\mathcal{N}_{\text{nr}}&=\frac{G_{\text{F}}\,\alpha_e\,}{2^7\,\pi^4\,m_D}\,\sqrt{\pi\,\frac{\sqrt{\lambda\,\lambda_p}}{m_D\,p^2}}\,,
\end{split}
\end{align}
where $\lambda=\lambda(m_D^2,\,q^2,\,p^2)$ and $\lambda_p=\lambda(p^2,\,m_{P_1}^2,\,m_{P_2}^2)$. 
In addition, 
\begin{align}
\begin{split}
    \omega_\pm&=\pm\frac{\hat g}{2}\,\frac{f_D}{f_{P_1}^2}\,\frac{m_D}{v\cdot p_{P_1}+\Delta}\,,\\
    h&=\frac{{\hat g}^2}{2}\,\frac{f_D}{f_{P_1}^2}\,\frac{1}{(v\cdot p_{P_1}+\Delta)\,(v\cdot p +\Delta)}\,,
\end{split}
\end{align}
with the decay constants $f_D$ and $f_{P_1}$~\cite{Tanabashi:2018oca}, $\Delta=(m_{D^{*0}}-m_{D^0})$,\,$\hat g=0.570\pm0.006$~\cite{Lees:2013zna} and the dot products
\begin{align}
    v\cdot p_{P_1}&=\frac{1}{4\,m_D}\,\Bigg((m_D^2-q^2+p^2) \notag\\ &-\sqrt{\lambda(m_D^2,\,q^2,\,p^2)\,\left(1-\frac{4\,m_{P_1}^2}{p^2}\right)}\,\cos{\theta_{P_1}}\Bigg)\,, \notag\\
    v\cdot p&=\frac{m_D^2-q^2+p^2}{2\,m_D}\,.
\end{align}
An isospin factors of $1/\sqrt2$ needs to be included into the form factors for each $\pi^0$ in the final state. 
Together with the statistical factor for identical particles, the $D^0 \to \pi^0\pi^0\nu\bar{\nu}$ mode receives an overall suppression by $1/2$ with respect to $D^0\to \pi^+\pi^- \nu\bar\nu$ in the isospin limit.

\subsection{Form factors $\boldsymbol{\Lambda_c^+ \to p}$}\label{app:ff_lambdac}
 
Taking into account the difference in notation between  Ref.~\cite{Meinel:2017ggx} and Section \ref{sec:baryons}, $f^V_{0,\perp} = f_{+,\perp}$ and $f^A_{0,\perp} = g_{+,\perp}$, the form factors $f_\perp$, $g_\perp$, $f_+$ and $g_+$ can be extracted from Ref.~\cite{Meinel:2017ggx}:
\begin{align}
f(q^2)=\frac{1}{1-q^2/(m^f_{\text{pole}})^2}\,\sum_{n=0}^2\,a_n^f\,[z(q^2)]^n,
\end{align}
with $z(q^2)$ given in Eq.~\eqref{eq:zexpansion}, 
$t_+ =(m_D + m_\pi)^2$ and $t_0 = (m_{\Lambda_c^+}-m_p)^2$.
The values for the $a_n$ parameters and the pole masses $m_{\text{pole}}$ are given along with their correlation in the supplemented files of Ref.~\cite{Meinel:2017ggx}.


\end{document}